\documentclass[aps,prd,12pt,notitlepage,showpacs,nofootinbib,tightenlines]{revtex4-1}
\usepackage{amsmath}
\usepackage{amssymb}
\usepackage{epsfig}
\usepackage{graphicx}
\usepackage{bm}
\usepackage{times}
\usepackage{braket}
\usepackage{color}
\usepackage{slashed}
\usepackage{hyperref}
\definecolor{Red}{rgb}{1.,0.,0.}

\definecolor{Blue}{rgb}{0.,0.,1.}

\definecolor{nicered}{rgb}{0.7,0.1,0.1}
\definecolor{nicegreen}{rgb}{0.1,0.5,0.1}
\hypersetup{colorlinks,citecolor=nicegreen,linkcolor=nicered}

\begin{document}
%%%%%%%%%%%%%%%%%%%%%%%%%%%%%%%%%%%%%%%%%%%%%

\newcommand{\beq}{\begin{eqnarray}}
\newcommand{\eeq}{\end{eqnarray}}
\newcommand{\non}{\nonumber\\ }
\newcommand{\ord}{{\cal O}}

\def\eeqn{\end{equation}}
\newcommand\iden{\leavevmode\hbox{\small1\normalsize\kern-.33em1}}

%%---------------------------------------
\def \cpc{ {\bf Chin. Phys. C} }
\def \csb{ {\bf Chin. Sci. Bull.} }
\def \ctp{ {\bf Commun.Theor.Phys. } }
\def \epjc{{\bf Eur.Phys.J. C} }
\def \jpg{ {\bf J.Phys. G} }
\def \npb{ {\bf Nucl.Phys. B} }
\def \plb{ {\bf Phys.Lett. B} }
\def \pr{  {\bf Phys. Rep.} }
\def \prd{ {\bf Phys.Rev. D} }
\def \prl{ {\bf Phys.Rev.Lett.}  }
\def \epl{ {\bf Europhys.Lett.}  }
\def \cpl{ {\bf Chin. Phys. Lett.}  }
\def \ptp{ {\bf Prog. Theor. Phys. }  }
\def \rmp{ {\bf Rev.Mod.Phys. }  }
\def \zpc{ {\bf Z.Phys.C}  }
\def \jhep{ {\bf J. High Energy Phys.}  }
\def \ijmpa{ {\bf Int. J. Mod. Phys. A}  }
\def \mpla{ {\bf Mod. Phys. Lett. A}  }
\def \scg{ {\bf Sci China Phys Mech $\&$ Astron}  }

\def\eslash{\rlap{\hspace{0.02cm}/}{E_{T}}}
\def\btt#1{{tt$\backslash$#1}}
\def\BibTeX{\rm B{\sc ib}\TeX}
\def\ov{ \overline }
%%---------------------------------------------------------

%%%%%%%%%%%%%%%%%%%%%%%%%%%%%%%%%%%%%%%%%%%%%%%%%%%%
%\begin{document}
%%
\title{The production and decay of the top partner $T$ in the left-right twin higgs model
at the ILC and CLIC }
\author{Yao-Bei Liu$^{1,2}$} \email{liuyaobei@sina.com}
\author{Zhen-Jun Xiao$^{1,3}$}\email{xiaozhenjun@njnu.edu.cn}
\affiliation{1. Department of Physics and Institute of Theoretical Physics,
                 Nanjing Normal University, Nanjing 210023, P.R.China }
\affiliation{2. Henan Institute of Science and Technology, Xinxiang 453003, P.R.China}
\affiliation{3. Jiangsu Key Laboratory for Numerical Simulation of Large Scale
Complex Systems, Nanjing Normal University, Nanjing 210023, P.R. China}
\date{\today}
\begin{abstract}
The left-right twin Higgs model (LRTHM) predicts the existence of the top partner $T$.
In this work, we make a systematic investigation for the single and pair production of
this top partner $T$ through the processes: $e^{+}e^{-}\to t\ov{T} + T\bar{t}$ and $ T\ov{T}$,
the neutral scalar (the SM-like Higgs boson $h$ or neutral pseudoscalar
boson $\phi^{0}$) associate productions $e^{+}e^{-}\to  t\ov{T}h +T\bar{t}h$,
$T\ov{T}h$, $t\ov{T}\phi^{0}+T\bar{t}\phi^{0}$ and $ T\ov{T}\phi^{0}$.
From the numerical evaluations for the production cross sections and relevant phenomenological
analysis we find that (a) the production rates of these processes, in the reasonable parameter space,
can reach the level of several or tens of fb;
(b) for some cases, the peak value of the resonance production cross section can be enhanced significantly
and reaches to the level of pb;
(c) the subsequent decay of $T\to \phi^{+}b \to t\bar{b}b$ may generate typical phenomenological
features rather different from the signals from other new physics models beyond the standard model(SM);
and (d) since the relevant SM background is generally not large, some signals of the top partner $T$
predicted by the LRTHM may be detectable  in the future ILC and CLIC experiments.
\end{abstract}


\maketitle

%%====================================================================
%%\newpage
\section{Introduction}

With the observation of a standard model (SM) Higgs
boson with a mass around 125 GeV \cite{atlas,cms,lhcphysics} at the Large Hadron Collider (LHC),
our understanding of electroweak symmetry breaking (EWSB) has been
significantly improved than before \cite{lhcphysics}.
However, this does not necessarily mean that the SM is fundamentally the whole story
\cite{zhu}.
It is well known that the SM has a serious problem called the little hierarchy
problem \cite{0007265}.
The twin Higgs mechanism \cite{twin1,twin2} has
been proposed recently to tackle this little hierarchy problem, in which the SM-like
Higgs emerges as a pseudo-Goldstone boson once a global
symmetry is spontaneously broken.
The twin Higgs theories use a discrete symmetry in combination with an approximate global symmetry
to eliminate one-loop quadratic divergence and thus stabilizing the mass of Higgs boson.

The twin Higgs mechanism can be implemented in left-right models with the additional discrete
symmetry being identified with left-right symmetry \cite{ly}.
The left-right twin Higgs model (LRTHM)
is a concrete realization of the twin Higgs mechanism \cite{Hock}. In this model,
the SM gauge symmetry is extended to $SU(2)_{L}\times SU(2)_{R}\times
U(1)_{B-L}$, which is embedded into the global $U(4)_{1}\times
U(4)_{2}$ symmetry.
The leading quadratically
divergent contributions of the SM gauge bosons to the Higgs boson
mass are canceled by the loop involving the new heavy gauge bosons ($W^{\pm}_{H}, Z_{H}$), while
those for the top quark can be canceled by the contributions from a
heavy top partner ($T$). These new particles predicted by the LRTHM at or below the TeV scale, which might generate
characteristic signatures at the present and future high energy colliders \cite{Hock,twin2,dong,lei,liu,ycx}.
Very recently, we have studied the properties of the LRTHM confronted with the latest LHC Higgs data \cite{liuprd}.

Recently, many searches have been performed by both ATLAS \cite{atlas-1,atlas-2} and CMS \cite{cms-1,cms-2} collaborations
in order to discover or set bounds on the heavy top-quark partner, assuming decays into three channels,
 $W^{+}b$, $Zt$ and $ht$, and scanning over various combinations of the
branching ratios. For instance, top partner
with masse below $656$ GeV are excluded at $95\%$ confidence level under the assumption of a branching
ratio $BR(T\rightarrow W^{+}b)=1$ \cite{atlas-3}. However, the dominant decay mode for the
top partner in the LRTHM is into a charged Higgs boson and a bottom quark. Thus, the current
bound on the top partner will be relaxed. The production
of the $T$-quark at the LHC have been described in Ref.
\cite{Hock}, in which the $s$-channel on shell $W_{H}$
decay dominated the single heavy top production. The single production of the top partner via
the $e\gamma$  and $\gamma\gamma$ fusion processes
has been studied in Refs. \cite{st1,st2}.

So far, most of the works about the top partner focus on phenomenological
analysis at the LHC experiments, see for example Refs.~\cite{t-lhc1,t-lhc2,t-lhc3}.
When compared with the LHC, a TeV scale linear $e^{+}e^{-}$ collider
has a particularly clear background environment, with a center of mass(c.m.) energy in the
range of 500 to 1600 GeV,  as in the case of the International Linear Collider(ILC) \cite{ILC}, and
 of 3 TeV to the Compact Linear Collider(CLIC) \cite{CLIC}.
The high luminosity linear collider is thus a precision machine with which the properties
of new particles can be measured precisely.
For example, the final stage of CLIC operating at an energy of 3 TeV is expected to
directly examine the pair production of new heavy top partner of mass up to 1.5 TeV \cite{CLIC1}.
A detailed study of the anomalous single fourth generation $t'$ quark  production
at ILC and CLIC has been performed in ref.~\cite{npb-851-289}.
The phenomenology of top partners in the little Higgs models with T-parity (LHT) and the
minimal supersymmetric standard model with R-parity (MSSMR) at future linear colliders
are studied in Refs.~\cite{t-ilc1,t-ilc2}, in which the decay signal of $T$-quark
($T\to tA_{H}$) can fake the signal of the scalar top quark
$\tilde{t}\rightarrow t\tilde{\chi}_{0}^{1}$.
In the LRTHM, furthermore,  the dominant decay mode $T\rightarrow\phi^{+}b \rightarrow t\bar{b}b$
may generate different phenomenological features. Thus, in this paper,
we will perform a comprehensive analysis on six top partner production processes:
$e^{+}e^{-}\rightarrow t\ov{T}+T\bar{t}, T\ov{T}, t\ov{T}h+T\bar{t}h, T\ov{T}h,
t\ov{T}\phi^{0}+T\bar{t}\phi^{0}$ and $e^{+}e^{-}\rightarrow T\ov{T}\phi^{0}$ at the future
possible ILC and/or  CLIC experiments.

This paper is organized as follows. In section II, we give a
brief review of the LRTHM, and then study the decays of the top partner
and the charged Higgs bosons. Sec. III is devoted to the computation of the production
cross section (CS) of above mentioned six production channels.
Some phenomenological analysis are also included in these three sections.
Our conclusions are given in section IV.

\section{Overview of the LRTHM}

The details of the LRTHM and some phenomenology analysis have been studied in Ref. \cite{Hock}. Thus we
will focus on the top partner sector in this section.
In the LRTHM, two Higgs fields ($H$ and $\hat{H}$) are introduced and each
transforms as $(4,1)$ and $(1,4)$ respectively under the global
symmetry. They are written as
\begin{eqnarray}
H=\left( \begin{array}{c} H_{L}\\ H_{R} \\
\end{array}  \right)\,,~~~~~~~~~~~~~~\hat{H}=\left( \begin{array}{c} \hat{H}_{L}\\ \hat{H}_{R} \\
\end{array}  \right)\,,
\end{eqnarray}
where $H_{L,R}$ and $\hat{H}_{L,R}$ are two component objects which
are charged under the $SU(2)_{L}\times SU(2)_{R}\times U(1)_{B-L}$
as
\begin{equation}
H_{L}~and~ \hat{H}_{L}: (2, 1, 1),~~~~~~~~H_{R}~ and~ \hat{H}_{R}:(1, 2, 1).
\end{equation}
The global $U(4)_{1}(U(4)_{2})$ symmetry is spontaneously broken
down to its subgroup $U(3)_{1}(U(3)_{2})$ with non-zero vacuum
expectation values (VEV) as $\langle H\rangle=(0,0,0,f)$ and $\langle
\hat{H}\rangle=(0,0,0,\hat{f})$. Each spontaneously symmetry
breaking yields seven Nambu-Goldstone bosons.
The gauge symmetry $SU(2)_{L}\times SU(2)_{R}\times U(1)_{B-L}$ is eventually broken down to the SM $U(1)_{em}$, six out of the 14 Goldstone bosons are eaten by the SM gauge bosons $(W^{\pm},Z)$
and the heavy gauge bosons $(W_{H}^{\pm},Z_{H})$ in the LRTH model.
After the re-parametrization of the fields,  the remaining 8 particles include one SM-like Higgs
boson $h$, one neutral pseudoscalar $\phi^{0}$, a pair of charged scalar $\phi^{\pm}$ and an extra $SU(2)_{L}$ doublet $\hat{h}=(\hat{h}_{1}^{+},\hat{h}_{2}^{0})$.
The lightest particle in the odd $\hat{h}_{2}^{0}$ is stable, and thus can be a candidate for dark matter.

The masses of the heavy gauge bosons are expressed as:
\begin{eqnarray}
M_{W_{H}}^{2}&=& \frac{1}{2}g^{2}(\hat{f}^{2}+f^{2}\cos^{2}x),\\
M_{Z_{H}}^{2}&=&
\frac{g^{2}+g'^{2}}{g^{2}}(M_{W}^{2}+M_{W_{H}}^{2})-M_{Z}^{2},\end{eqnarray}
where $x=v/(\sqrt{2}f)$ and $v$ is the electroweak scale, the values
of $f$ and $\hat{f}$ are interconnected once
we set $v=246$ GeV. The Weinberg angle can be written as:
\begin{eqnarray}
s_{W}=\sin\theta_{W}=\frac{g'}{\sqrt{g^{2}+2g'^{2}}},~~~~~~~~ c_{W}=\cos\theta_{W}=\sqrt{\frac{g^{2}+g'^{2}}{g^{2}+2g'^{2}}}.
\end{eqnarray}

%\textcolor{red}{}

Besides the SM-like Higgs boson $h$, both the charged scalars $\phi^{\pm}$ and the neutral pseudoscalar $\phi^{0}$ can couple
to both the fermions and the gauge bosons. Their masses can be obtained from the one-loop Coleman-Weinberg (CW) potential and the soft left-right symmetry breaking terms, so-called $\mu-$term \cite{Hock}:
\beq
V_{\mu}=-\mu_{r}^{2}(H_{R}^{\dagger}\hat{H}_{R}+h.c.)+\hat{\mu}^{2}H_{L}^{\dagger}\hat{H}_{L}.
\eeq
Here $\hat \mu$ is of the order of $f$ or smaller, and $\mu_r$ should be less than
about $f/4\pi$ in order not to reintroduce fine tuning \cite{Hock}.
The masses of  $\phi^{0}$ and $\phi^{\pm}$ can therefore be written as the form of
\begin{eqnarray}
m^{2}_{\phi^0}&=&\frac{\mu_{r}^{2}f\hat{f}}{\hat{f}^{2}+f^{2}\cos^{2}x}
 \cdot \left \{\frac{\hat{f}^{2}\left[ \cos
x+\frac{\sin x}{x}(3+x^{2}) \right]}{f^{2}\left (\cos x+\frac{\sin x}{x} \right )^{2}}+2\cos
x+\frac{f^{2}\cos^{2}x(1+\cos x)}{2\hat{f}^{2}} \right\},\label{eq:mphi2}\\
m^{2}_{\phi^\pm}&=&\frac{3}{16\pi^2}\frac{g'^{2}M^2_{W_H}}{M^2_{Z_H}-M^2_{Z}}\Big[(\frac{M^2_{W}}{M^2_{Z_H}}-1){\mathcal Z}(M_{Z_H})
-(\frac{M^2_{W}}{M^2_{Z}}-1){\mathcal Z}(M_{Z})\Big]\nonumber\\
&&\,+\frac{\mu_{r}^{2}f\hat{f}}{\hat{f}^{2}+f^{2}\cos^{2}x}(\frac{\hat{f}^{2}x}{f^{2}\sin x}+2\cos
x+\frac{f^{2}\cos^{3}x}{\hat{f}^{2}}),
\end{eqnarray}
where ${\mathcal Z}(x)=-x^2(\ln\frac{\Lambda^2}{x^2}+1)$, and the
cut-off scale $\Lambda$ is typically taken to be $4 \pi f$. In the allowed parameters space, the masses of the charged scalars  $\phi^{\pm}$ are generally in the range of a few hundred GeV. The value of
$m_{\phi^0}^2$ depend on two parameters $\mu_r$ and $f$, and the lower limit comes from the
non-observation of the decay $\Upsilon\to \gamma + X_{0}$ \cite{cons-phi0}.

\subsection{Masses and relevant couplings of top quark sector}

In order to cancel the one-loop quadratic
divergence
 of Higgs mass induced by top quark, a pair
of vector-like quarks ($U_{L}, U_{R}$) are introduced, which are singlets under
$SU(2)_{L}\times SU(2)_{R}$. The Lagrangian can be written as \cite{Hock}
\begin{equation}
{\cal L}_{t}=y_L\bar{Q}_{L3}\tau_2 H_L^*U_R
  +y_R\bar{Q}_{R3}\tau_2H_R^*U_L - M\bar{U}_LU_R + h.c.
\label{yukawatop}
\end{equation}
where $Q_{L3} = -i(u_{L3},d_{L3})^T$ and $Q_{R3} =
(u_{R3},d_{R3})^T$. Under left-right symmetry, $y_L=y_R=y$.
The mass eigenstates for the top quark $t$ and heavy top partner $T$ can be obtained
by diagonalizing the mass matrix.
Their masses and relevant couplings to gauge bosons are given by \cite{Hock}
\begin{eqnarray}
m_{t}^{2}&=&\frac{1}{2}(M^{2}+y^{2}f^{2}-N_{t}),~~~~~~~~ m_{T}^{2}=
\frac{1}{2}(M^{2}+y^{2}f^{2}+N_{t}),\\
Z_{\mu}t\bar{T}&:&e\gamma_{\mu}(C_{L}S_{L}\hat{f}^{2}c^{2}_{W}P_{L}+f^{2}x^{2}s^{2}_{W}C_{R}S_{R}P_{R}/(2\hat{f}^{2}c^{3}_{W});\\
Z_{H\mu}t\bar{T}&:&e\gamma_{\mu}(C_{L}S_{L}s^{2}_{W}P_{L}-c^{2}_{W}C_{R}S_{R}P_{R}/(2s_{W}c_{W}c2_{W});\\
Z_{\mu}T\bar{T}&:&-e\gamma_{\mu}(4s^{2}_{W}-3S^{2}_{L}P_{L})/(6s_{W}c_{W});\\
Z_{H\mu}T\bar{T}&:&-e\gamma_{\mu}[(3C^{2}_{L}+1)s^{2}_{W}P_{L}-(3c^{2}_{W}C^{2}_{R}-4s^{2}_{W})P_{R}]/(6s_{W}c_{W}c2_{W});\\
Z_{H\mu}t\bar{t}&:&-e\gamma_{\mu}[(3S^{2}_{L}+1)s^{2}_{W}P_{L}-(3c^{2}_{W}S^{2}_{R}-4s^{2}_{W})P_{R}]/(6s_{W}c_{W}c2_{W}),
\end{eqnarray}
where
\begin{eqnarray}
S_{L}&=&\frac{1}{\sqrt{2}}\sqrt{1-(y^{2}f^{2}\cos 2x+M^{2})/N_{t}},~~C_{L}=\sqrt{1-S^{2}_{L}},\\
S_{R}&=&\frac{1}{\sqrt{2}}\sqrt{1-(y^{2}f^{2}\cos 2x-M^{2})/N_{t}},~~C_{R}=\sqrt{1-S^{2}_{R}},
\end{eqnarray}
with $N_{t}=\sqrt{(M^{2}+y^{2}f^{2})^{2}-y^{4}f^{4}\sin^{2}2x}$ and $x=v/(\sqrt{2}f)$.

%%----------------------------------------------
\begin{figure}[ht]
\begin{center}
\scalebox{0.6}{\epsfig{file=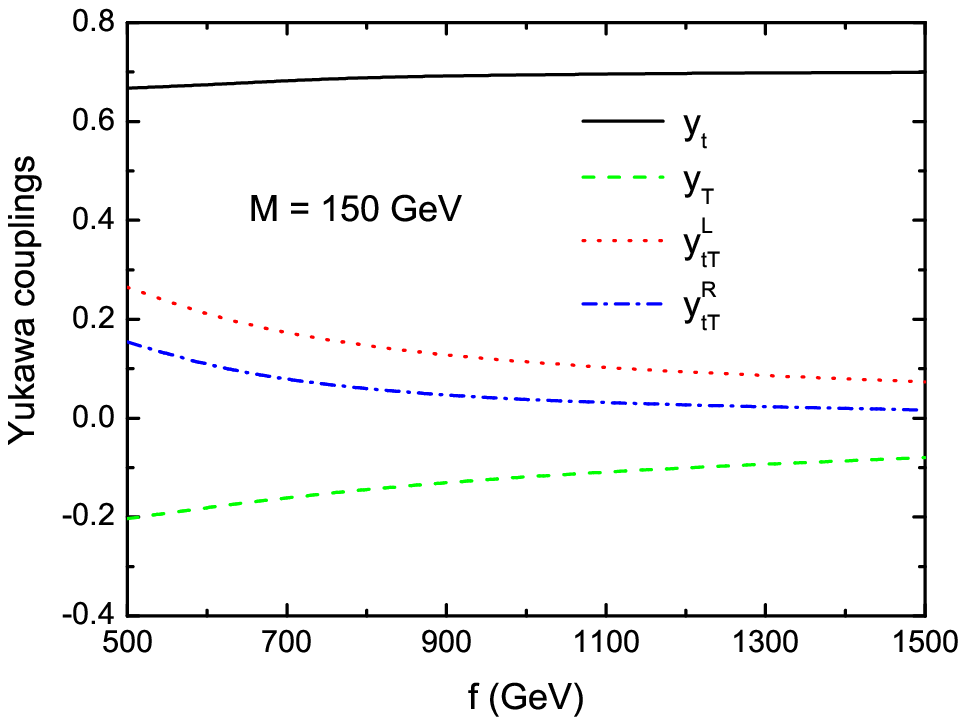}\epsfig{file=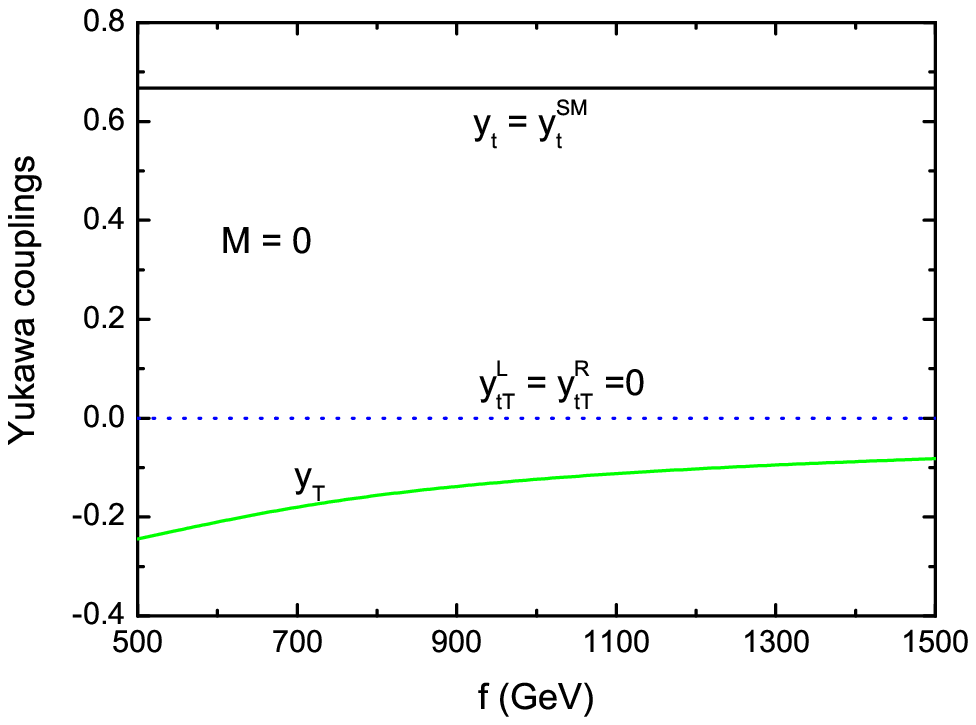}}
\end{center}
\caption{ The Yukawa couplings ($y_{t}, y_{T}, y^{L}_{tT}, y^{R}_{tT}$) as a function of
the parameter $f$ for two typical values of $M$.}
\end{figure}
%%----------------------------------------------

From Eq.(\ref{yukawatop}), we can get the interactions between the Higgs boson
and the pairs of $(t\bar{t},T\ov{T},T\bar{t},\ov{T}t)$:
\beq
{\mathcal{L}_{int}}=-y_{t}t\bar{t}h-y_{T}T\bar{T}h-(\bar{T}[y^{L}_{tT}P_{L}+y^{R}_{tT}P_{R}]th+h.c.),
\eeq
where the Yukawa couplings constants ($y_{t}, y_{T}, y^{L}_{tT}, y^{R}_{tT}$) are defined as
\begin{eqnarray}
y_{t}&=&-\frac{m_{t}}{v}C_{L}C_{R},~~~~~~~~~~~~~~~~~~~~~y_{T}=-y(S_{R}S_{L}-C_{L}C_{R}x)/\sqrt{2},\\
y^{L}_{tT}&=&-\frac{y}{\sqrt{2}}(C_{L}S_{R}+S_{L}C_{R}x),~~~~~~y^{R}_{tT}=-\frac{y}{\sqrt{2}}(C_{L}S_{R}x+S_{L}C_{R}).
\end{eqnarray}

Since the mixing angles are sensitive to the parameters $M$ and $f$, we plot in Fig.~1 the coupling constants of
the Yukawa interactions ($y_{t}, y_{T}, y^{L}_{tT}, y^{R}_{tT}$) as a function of
the parameter $f$ for two typical values of $M$: $M=0,150$ GeV. For $M=150$ GeV, the left-handed mixing
of top quark and top partner is larger than that for the right-handed ones, while
they all equal zero for $M=0$. In this case the top quark is purely $(u_{3L}, q_{R})$ and the top partner is purely $(q_{L}, u_{3R})$. On the other hand,
we can see that the couplings $y_{T}$ and $y_{t}$ have different sign, and $y_{t}$ is almost three times as large as $y_T$.

\subsection{Decays of the top partner and charged Higgs bosons}

In the LRTHM, the top quark partner $T$ can decay into
$\phi^{+}b$, $ht$, $Zt$, $Wb$ and $\phi^{0}t$, among which the decay
$T\rightarrow \phi^{+}b$ is the most important channel. Here we take the mass of the
charged scalars as $m_{\phi^{\pm}}=200$ GeV.
In Fig.~2 we show the $M$- and $f$-dependence of the branching ratios of
those relevant decays of the top quark partner $T$.
As shown in Fig.~2a, more than $60\%$ of top partner decays via
$T\rightarrow \phi^{+}b$ for 500 GeV$\leq f\leq$1000 GeV.
Other decays are strongly suppressed since the relevant couplings are suppressed by the
ratio $(M/f)$.
For comparison, the subdominant decay $T\rightarrow Wb$ can have a
branching ratio of about $11\%$ for $M=150$ GeV and $f$=500 GeV.
This is different from the case of the little Higgs model with T-parity, where the decay
$T\rightarrow W^{+}b$ is the dominant channel \cite{lht}.
In the limit $M=0$, the only two body decay mode is $T\rightarrow
\phi^{+}b$, with a branching ratio of $100\%$.
Thus, the previous bounds on the top partner will be relaxed in the LRTHM.
\begin{figure}[ht]
\begin{center}
\vspace{-0.5cm}
\scalebox{0.42}{\epsfig{file=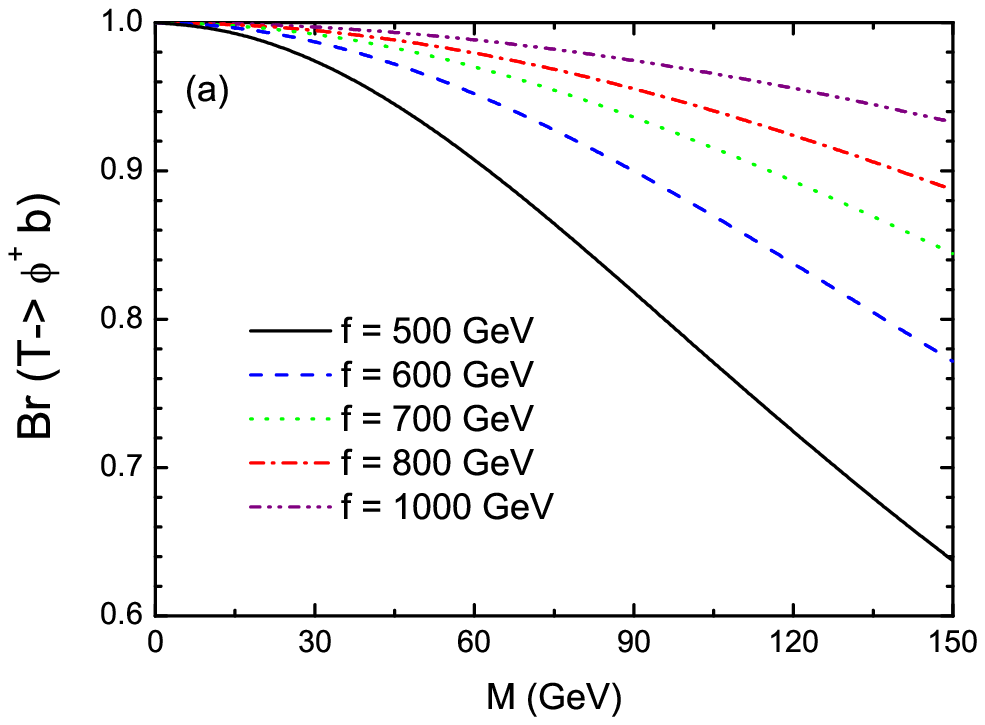} \epsfig{file=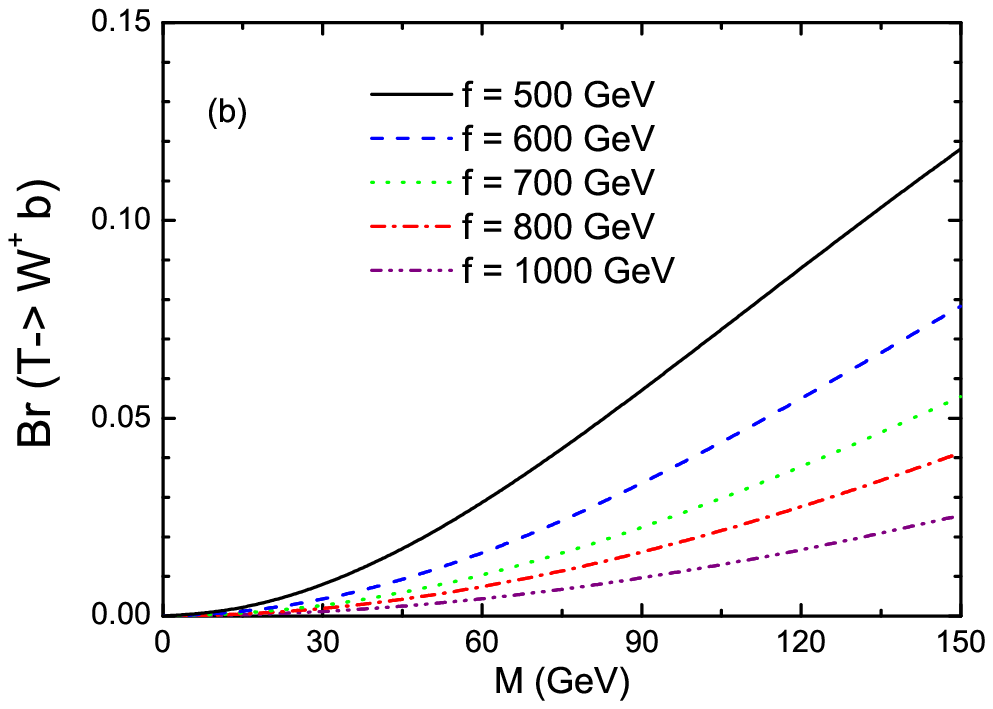} \epsfig{file=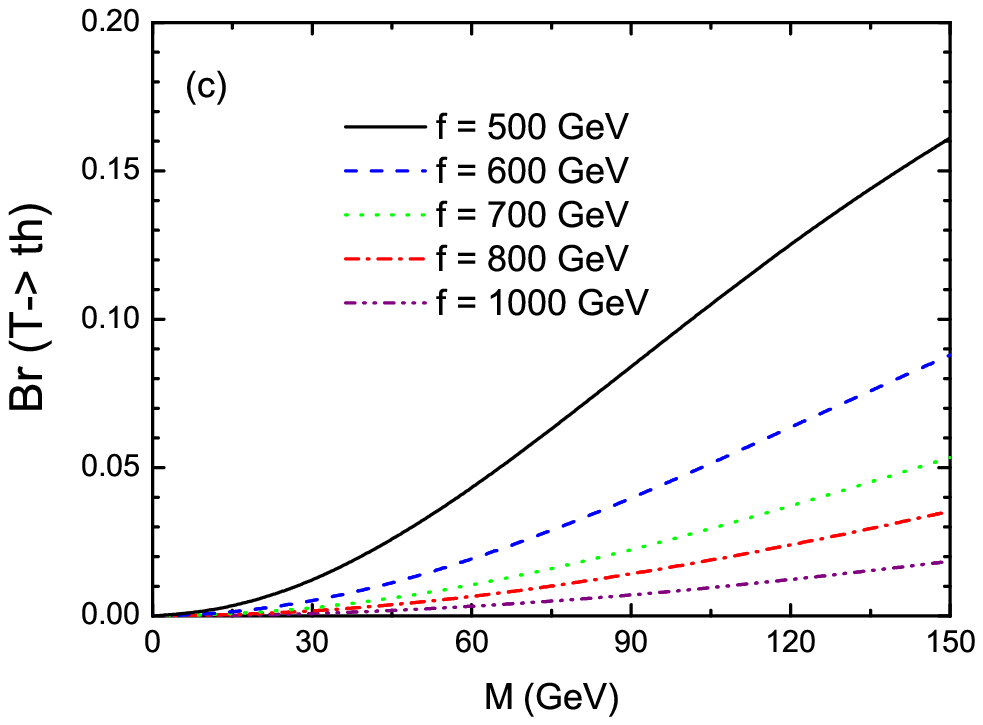}}
\scalebox{0.42}{\epsfig{file=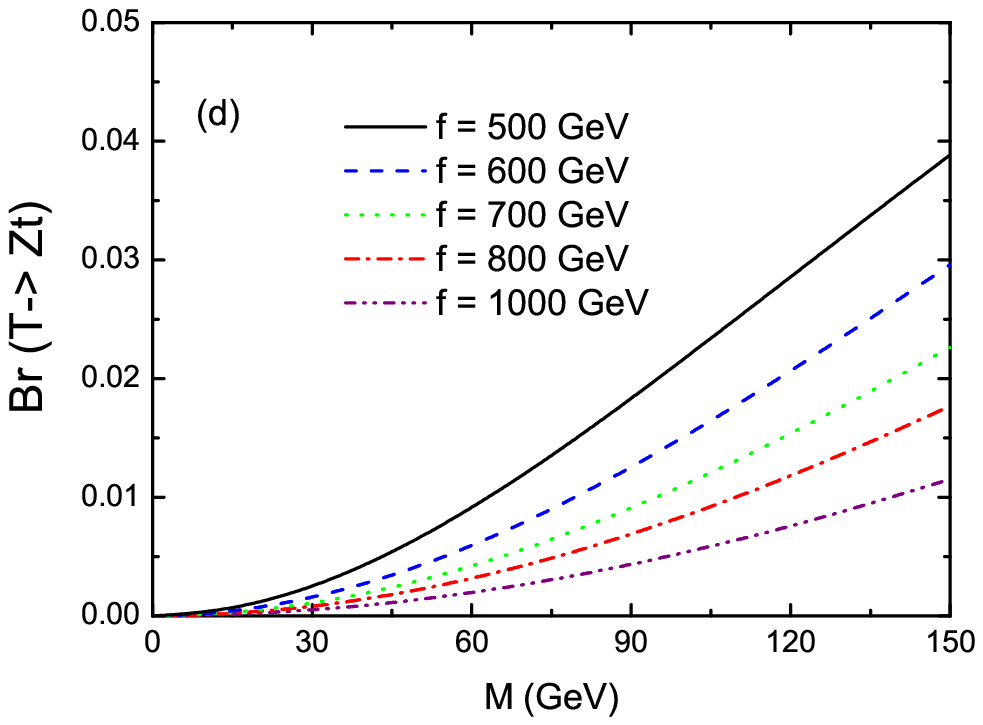} \epsfig{file=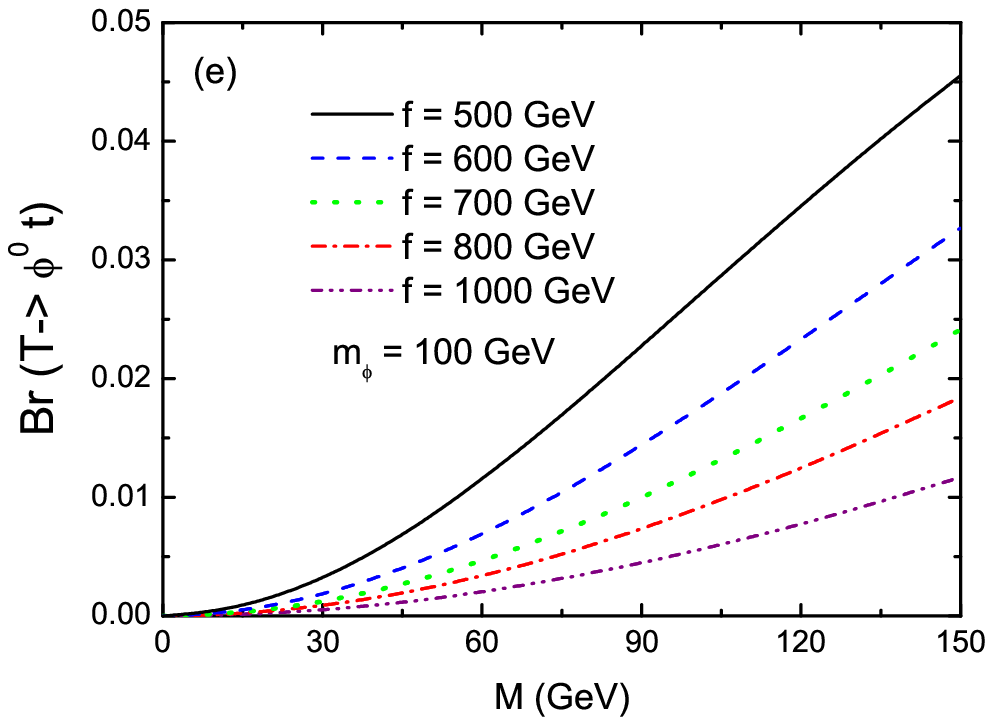} \epsfig{file=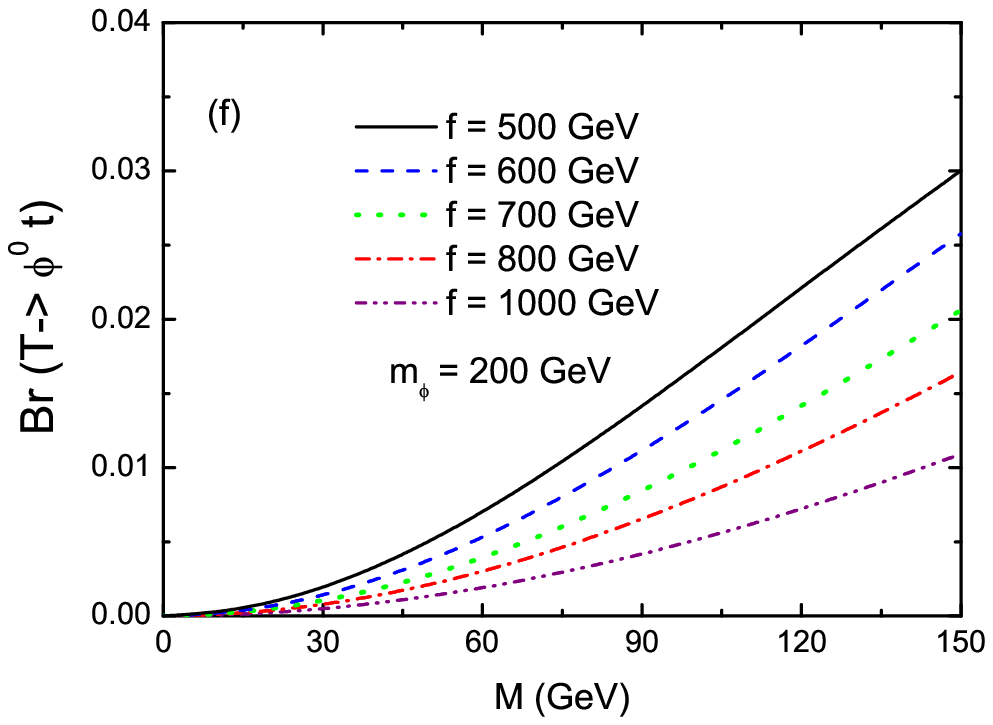}}
\end{center}
\vspace{-1cm}
\caption{ The branching ratios for various $T$ decay modes
as a function of the mixing parameter $M$ for five typical values of $f$.}
\label{fig:fig2}
\end{figure}

\begin{figure}[ht]
\begin{center}
\scalebox{0.75}{\epsfig{file=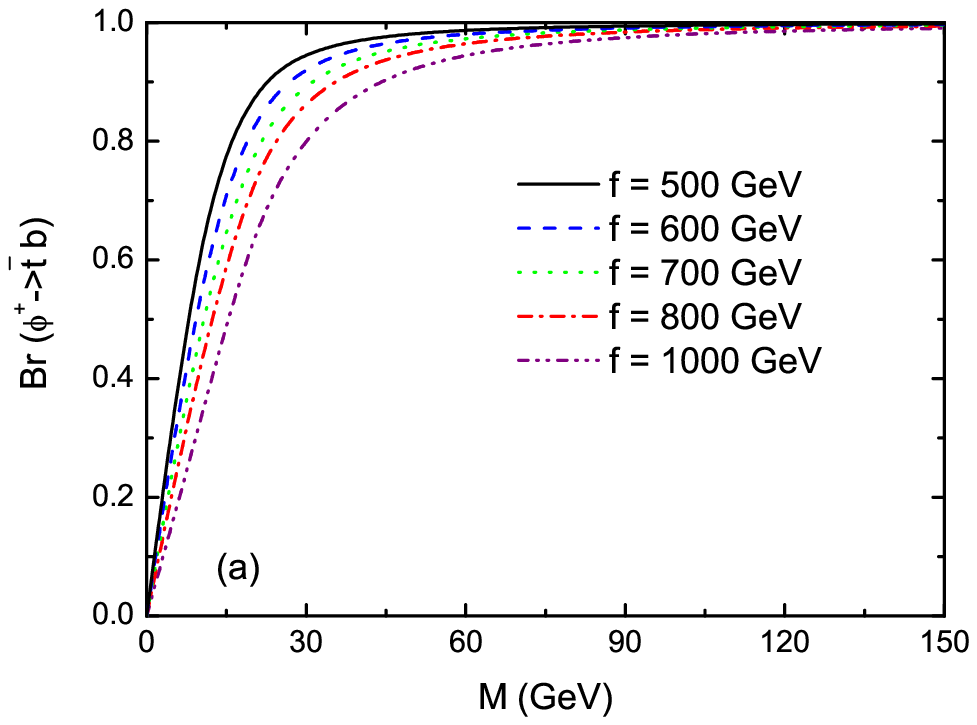}\epsfig{file=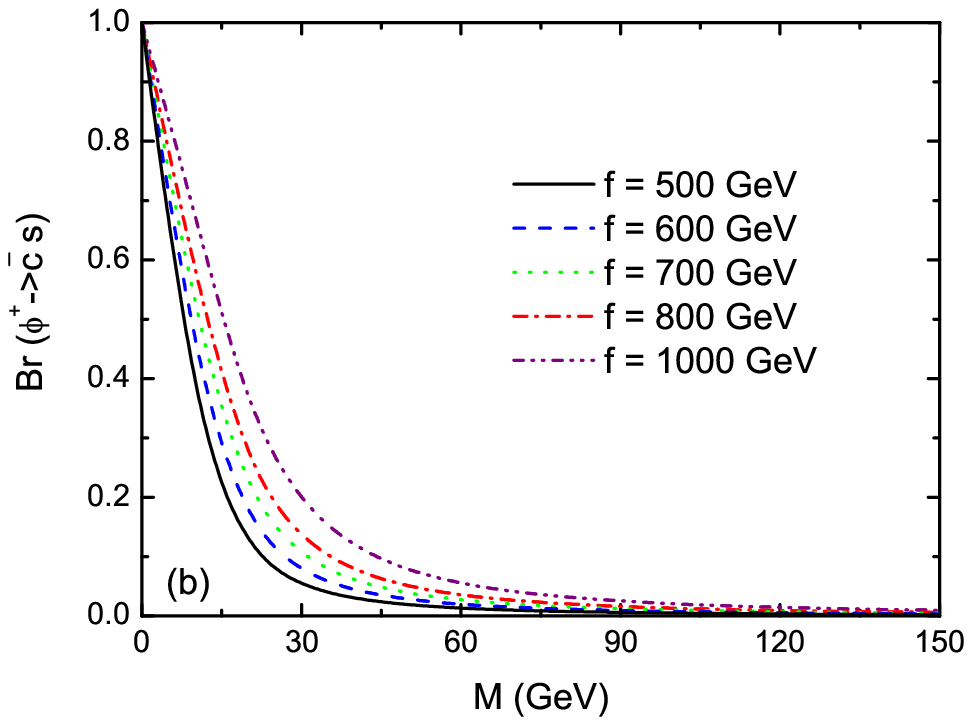}}
\end{center}
\caption{ The theoretical predictions for the $M$-dependence of the branching ratios
of $\phi^{+} \to t\bar{b}$ and $\phi^+ \to c\bar{s}$ decays, assuming five typical values
of $f$.} \label{fig:fig3}
\end{figure}

In the LRTHM, the charged Higgs $\phi^{\pm}$ decay  dominantly into quark pair $tb$ or $cs$ \cite{Hock}.
Fig.~3 shows the LRTHM predictions for the branching ratios for those two decay modes
as a function of the the mixing parameter $M$ for five typical values of parameter $f$.
One can see from Fig.~3 that the branching ratio of $\phi^{+}\rightarrow t\bar{b}$ decay
becomes larger than $50\%$ for large values of $M$.
While for very small values of $M$, $\phi^{+}\rightarrow c\bar{s}$ decay dominates, which may
lead to completely different phenomenology.
For $M=5$ GeV, for example, the branching ratio of $\phi^{+}\rightarrow c\bar{s}$ decay
will change from $65.2\%$ to $89.2\%$ when the parameter $f$
increases from 500 GeV to 1000 GeV.
In the lower limit $M=0$, the branching ratio of $T\rightarrow c\bar{s}$ is $100\%$.

\section{Numerical results and discussions}

The SM input parameters relevant in our study are taken as
$\alpha_{e}=1/128.8$, $S^{2}_{W}=0.2315$, $m_{Z}$=91.187 GeV
\cite{pdg2012} and $m_{t}$=173.3 GeV \cite{topmass}. The free LRTHM
parameters are $f$ and $M$. Note that the top Yukawa coupling $y$ can be determined by fitting the experimental value of the
top quark mass. The masses of top partner and heavy neutral gauge boson
can be determined by the value of $f$ and $M$. The typical
values of the top partner mass $m_T$, the heavy neutral gauge bosons masse $m_{Z_H}$ and decay
width $\Gamma_{Z_H}$ are listed in Table \ref{tab:table1} for several benchmark points
of the parameter $f$: $f=500,600,700,800,900,1000,1200$ and $f=1500$ GeV.

\begin{table}[ht]
\begin{center}
\caption{ The masses (in GeV) of the top partner $T$, the heavy neutral gauge boson $Z_H$ and the
total decay width $\Gamma_{Z_{H}}$ used in this paper, assuming $100\leq f \leq 1500$ GeV. }
\label{tab:table1}
\vspace{0.1in}
\doublerulesep 0.8pt \tabcolsep 0.1in
\begin{tabular}{|l|l|l|l|l|l|l|l|l|}\hline
       $f$ (GeV) &500&600&700&800&900&1000&1200&1500\\
\hline $m_{T}(M=0)$&466.4&571.3&674.5&776.8&878.4&979.5&1181&1482\\
\hline $m_{T}(M=150)$&489.9&590.7&691&791.1&891&991&1190.5&1489.5\\
\hline $m_{Z_{H}}$&1403&1684&1966&2247&2528&2810&3372&4215\\
\hline $\Gamma_{Z_{H}}$&29.8&35.7&41.6&47.4&53.3&59.2&71&88.7\\
\hline
\end{tabular} \end{center} \end{table}

In the LRTHM, the phenomenological studies on the signatures of the heavy neutral gauge boson $Z_H$
can be found in Ref.~\cite{lrth-z1}.
The present constraints on the  $Z'$ mass have been presented in \cite{pdg2012}.
The ATLAS and CMS experiments at the LHC have updated the
Tevatron limits on the heavy neutral gauge boson $Z'$ mass \cite{lhc-z}.
Recently, the ATLAS and CMS Collaborations presented results on narrow resonances with dilepton
final states and excluded a sequential standard model
$Z'$ with mass smaller than $2.49$ TeV \cite{atlas-z} and $2.59$ TeV \cite{cms-z}.
Based on the analysis of heavy resonances decaying into $t\bar{t}$ pairs with subsequent
fully hadronic and leptonic final states,
the ATLAS \cite{atlas-ztt} and CMS \cite{cms-ztt} collaborations also excluded the
leptophobic $Z'$ boson with the mass smaller than $1.32$ TeV (ATLAS) and $1.3$ TeV (CMS).
Using constraints from the precision electroweak (EW) data, the lower mass limit on extra
neutral boson $Z'$ in left-right symmetric models is around $1$ TeV \cite{lrsm}.
Although the Atlas and CMS data have been interpreted in terms of different scenarios for physics
beyond the SM, there is no any limit on $Z'$ in the LRTHM at present.
Our previous study using D0 and CDF results have excluded a $Z'$ in the LRTHM
with a mass below $940$ GeV \cite{lrth-z2}.

The indirect constraints on $f$ come
from the $Z$-pole precision measurements, the low energy neutral current process and high energy precision
measurements off the $Z$-pole, requiring approximately $f> 500$ GeV. On the other hand, it cannot be too
large since the fine tuning is more severe for larger $f$.
The value of the mixing parameter $M$ is constrained by the
$Z\to b\bar{b}$ branching ratio and oblique parameters.
Following Ref.~\cite{Hock}, we take the typical parameter space as:
\beq
500 GeV \leq f \leq 1500 GeV, \quad  0  \leq M \leq 150 GeV.
\eeq
All the numerical studies are done using CalcHEP \cite{calchep}.

\subsection{The single and pair production of top partner}

From above discussions, we know that the top partner can be singly or pair produced through
$s$-channel gauge bosons exchange by $e^{+}e^{-}$ collisions at ILC and CLIC energies.
The relevant Feynman diagrams are depicted in Fig.~4.
\begin{figure}[ht]
\begin{center}
\epsfig{file=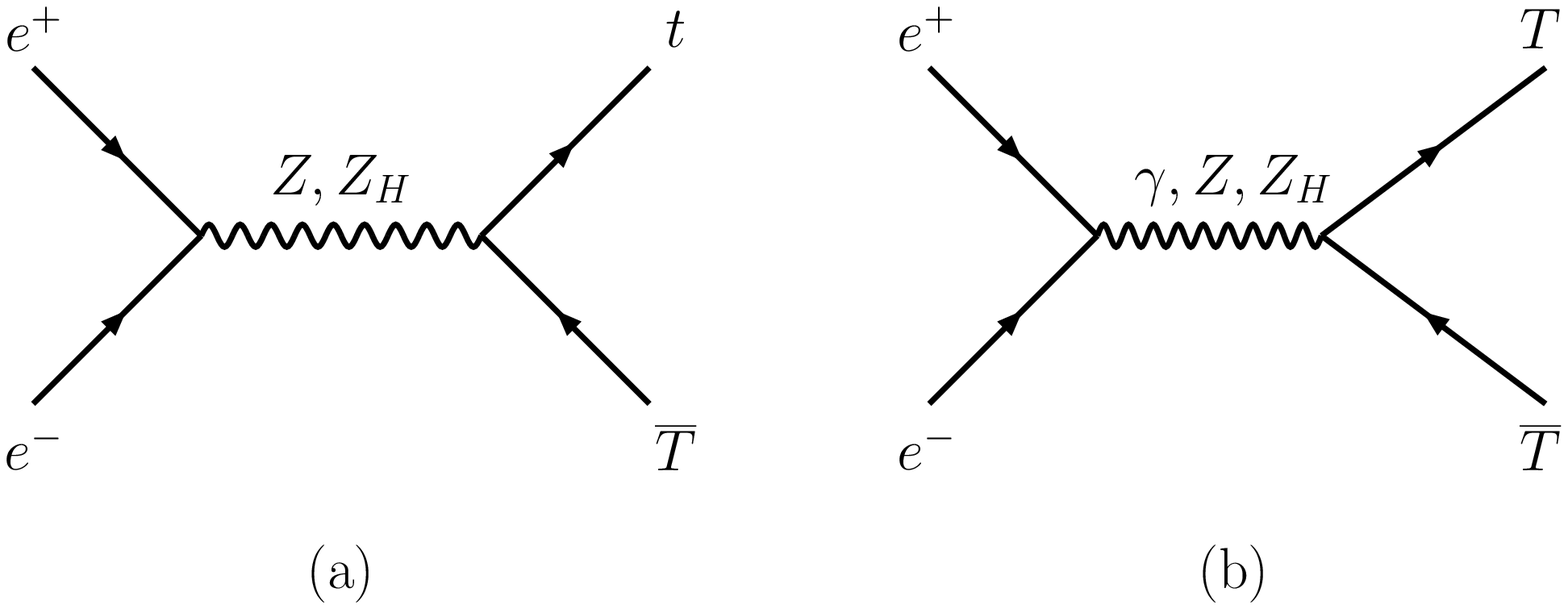,width=300pt,height=400pt}
\vspace{-9.5cm}
\caption{Feynman diagrams of the processes $e^{+}e^{-}\to t\ov{T}$ and $e^{+}e^{-}\to T\ov{T}$
in the LRTHM.}
\label{fig:fig4}
\end{center}
\end{figure}
\subsubsection{The $e^{+}e^{-}\rightarrow t\ov{T}+T\bar{t}$ process}

We fist consider the associate production of one top partner $T$ together
with the top quark through the $s$-channel $Z$ and $Z_H$ exchanges.
In Fig.~5a, we plot the production CS $\sigma (e^{+}e^{-}\to t\ov{T}+T\bar{t})$
as a function
of the mixing parameter $M$ for $\sqrt{s}=1.5$ TeV and five typical values of $f$.
One can see that the cross section decreases as the parameter $f$ increases. This
is natural since the phase space is depressed strongly by large
$m_{T}$. For $f=600$ GeV and $\sqrt{s}=1.5$ TeV, the maximum of the
cross section reaches the level of a few fb. On the other hand, the
results also show that the large $M$ can
enhance the cross section significantly. In the limit of $M=0$, its value goes to zero.

From Fig.~5b, one can see that the resonance peak of the cross
section $\sigma$ emerges when the $Z_{H}$ mass
$m_{Z_{H}}$ approaches the c.m. energy.
In the region of the resonance peak, the production CSs will be enhanced significantly
and can reach the order of pb.
For $\sqrt{s}=1.5$ TeV and $f$=700 GeV, for example, the value of $\sigma$ is about $5$ fb.
If we assume that the future ILC experiment with $\sqrt{s}$=1.5 TeV has a yearly integrated
luminosity of 500fb$^{-1}$, then there will be several thousand signal events
generated at the ILC.

\begin{figure}[th]
\begin{center}
\scalebox{0.75}{\epsfig{file=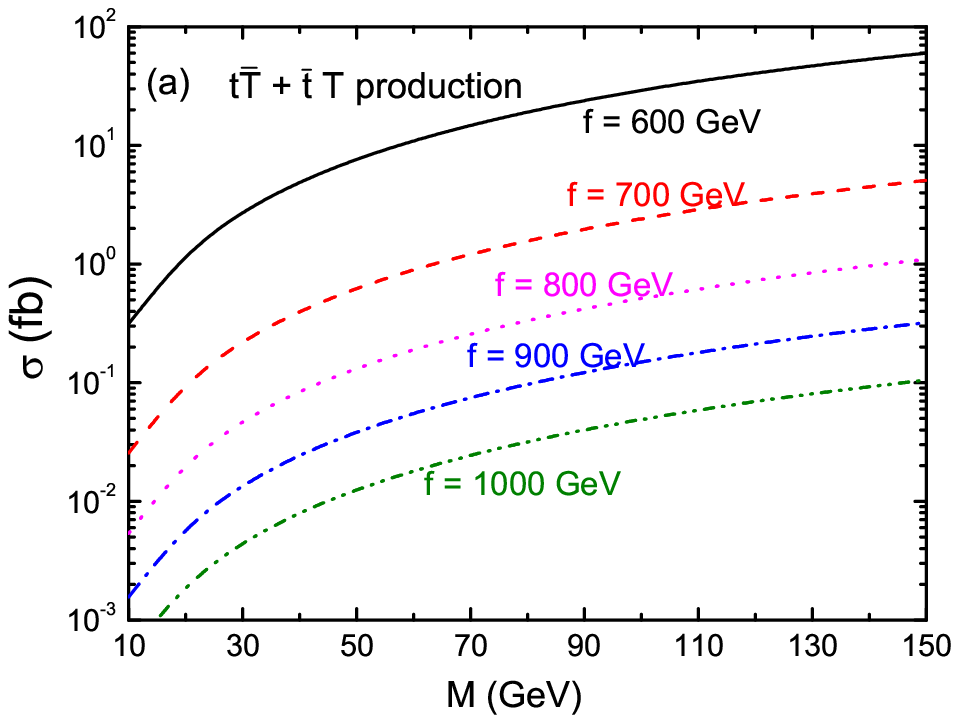}\epsfig{file=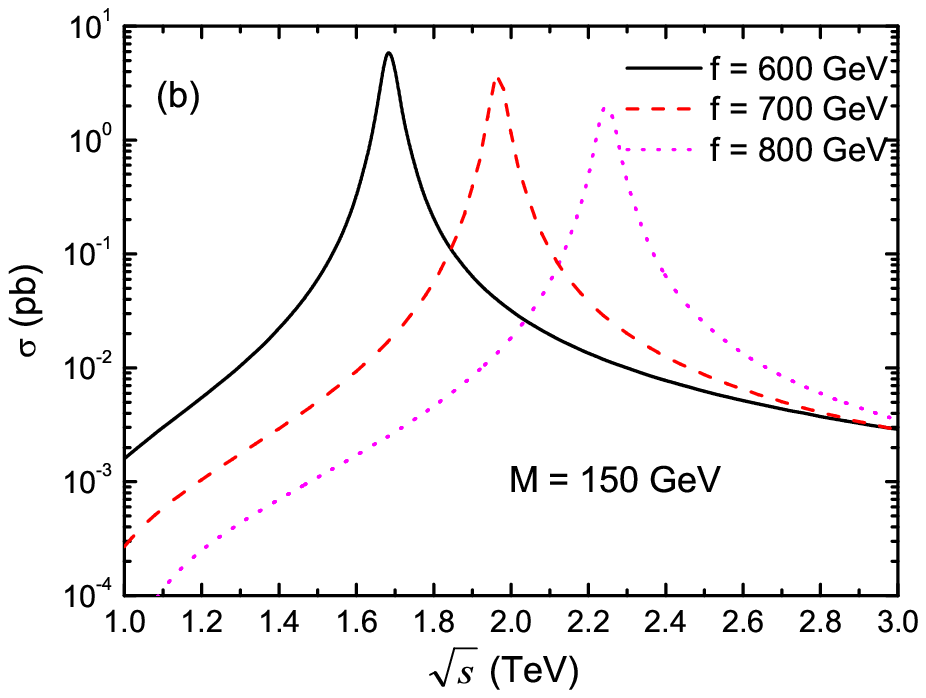}}
\end{center}
\caption{(a) The production CS $\sigma$ as a function
of the mixing parameter $M$ for $\sqrt{s}=1.5$ TeV and $f=600,
700, 800, 900$, and 1000 GeV; (b)The production CS $\sigma$ as a function
of center of mass energy $\sqrt{s}$ for $M=150$ GeV and three values of $f$ as indicated.}
\label{fig:fig5}
\end{figure}

For a large value of $M$, the dominate subsequent decay of $T\to \phi^{+}b$ and $\phi^{+}\to t\bar{b}$
make the process $e^{+}e^{-}\to t\ov{T}+T\bar{t}$  mainly decaying to the final
state $t\bar{t}b\bar{b}$.
The production rates for such final states can be easily estimated as
\beq
&&\sigma\times \Bigl[ Br(T\to \phi^{+}b)\cdot Br(\phi^{+}\to t\bar{b})
+Br(T\to th)\cdot Br(h\to b\bar{b})\non
&& \hspace{2cm}
+Br(T\to tZ)\cdot BR(Z\rightarrow b\bar{b})
+Br(T\rightarrow t\phi^{0})\cdot Br(\phi^{0}\rightarrow b\bar{b}) \Bigr].
\eeq
For the semi-leptonic decays of $t\bar{t}$, the characteristic collider signal
is two jet + four b + one lepton ($e$ or $\mu$) + missing $\eslash$.
The dominant SM background processes and their production CS's with
$\sqrt{s}=1.5$ TeV are listed in Table \ref{tab:table2}.
The backgrounds $t\bar{t}h$ and $t\bar{t}Z$ are also included
when $t\bar{t}b\bar{b}$ is estimated.  We can see that the total background
CS is about 0.4 fb. Note that these numerical results are
estimated by using MadGraph \cite{mad} and
cross-checked with CalcHEP without considering any kinematical cuts and tagging efficiency.

\begin{table}[]
\begin{center}
\caption{The possible SM background CS's (in fb) in semi-leptonic channel
($2j+4b+\ell+\eslash$) are estimated with $\sqrt{s}=1.5$ TeV. We used $Br(t\rightarrow W^{+}b)=1$,
$Br(W^{\pm}\rightarrow jj')=0.68$, $Br(W^{\pm}\rightarrow \ell^{\pm}\nu_{l})=0.107$,
$Br(h\rightarrow b\bar{b})=0.57$  and $Br(Z\rightarrow b\bar{b})=0.15$.} \label{tab:table2}
\vspace{0.1in}
\doublerulesep 0.8pt \tabcolsep 0.1in
\begin{tabular}{|l|l|}
\hline Processes &Cross sections (fb)\\
 \hline $e^{+}e^{-}\rightarrow t\bar{t}b\bar{b}$&$\sigma(e^{+}e^{-}\rightarrow t\bar{t}b\bar{b}\rightarrow 2j+4b+\ell+\eslash)=0.4$\\
\hline $e^{+}e^{-}\rightarrow W^{+}W^{-}ZZ$&$\sigma(e^{+}e^{-}\rightarrow W^{+}W^{-}ZZ\rightarrow 2j+4b+\ell+\eslash)=0.006$\\
\hline $e^{+}e^{-}\rightarrow W^{+}W^{-}hh$&$\sigma(e^{+}e^{-}\rightarrow W^{+}W^{-}hh\rightarrow 2j+4b+\ell+\eslash)=0.008$\\
\hline $e^{+}e^{-}\rightarrow W^{+}W^{-}Zh$&$\sigma(e^{+}e^{-}\rightarrow W^{+}W^{-}Zh\rightarrow 2j+4b+\ell+\eslash)=0.002$\\
\hline
 \end{tabular}
\end{center}\end{table}

%%=====================================
\begin{figure}[ht]
\begin{center}
\scalebox{0.85}{\epsfig{file=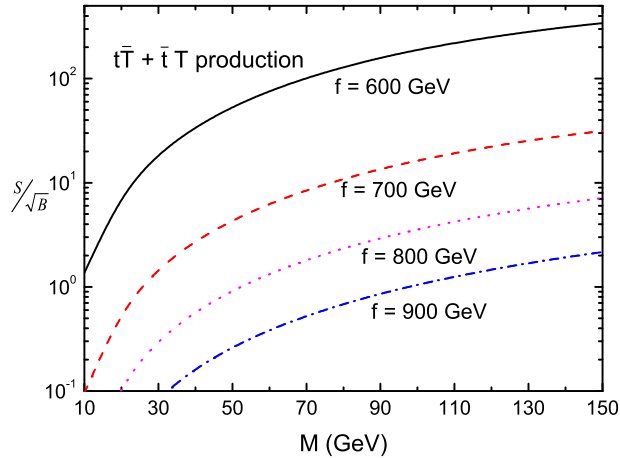}}
\end{center}
\caption{ The statistical significance $S/\sqrt{B}$ as a function of the the mixing
parameter $M$ for $\sqrt{s}=1.5$ TeV and four typical values of $f$.}
\end{figure}
%%=====================================

In order to discuss the observation of the top partner, we calculate the statistical significance
$S/\sqrt{B}$ ($S$ denotes the signal and $B$ the SM background) and the numerical results
are shown in Fig.~6, here we assumed that the integrated luminosity is $500$ fb$^{-1}$.
One can see that, for large $M$ and small $f$, the value of the statistical significance
$S/\sqrt{B}$ is larger than 5. For $f\geq 600$ GeV, the mass of the heavy gauge boson is
larger than 1680 GeV and the resonance peak will not appear.
Consequently, it may be possible to extract the signals from the backgrounds in
the reasonable parameter space of the LRTHM.

It is obvious that this is only a simple estimate. To take into account detector acceptance we should consider the
tagging efficiency and some appropriate kinematical cuts.
On the other hand, the reconstruction of the top partner and the charged Higgs bosons
is very necessary to distinguish the signal from the background.
In our estimates, we have excluded the efficiency
$\epsilon_{b}^{4}$ of tagging the four $b$-jets in the final state.
If we take the single $b$-tagging efficiency as about $70\%$, as one would expect,
after putting some basic acceptance cuts required to trigger on the
final states, the rates would become smaller.
However, our main conclusions should remain unchanged.
Obviously, the detailed analysis for individual processes would require Monte-Carlo
simulations of the signals and backgrounds, which is beyond the scope of the current paper.

\subsubsection{The $e^{+}e^{-}\to T\ov{T}$ process}

%%=====================================
\begin{figure}[th]
\begin{center}
\scalebox{0.8}{\epsfig{file=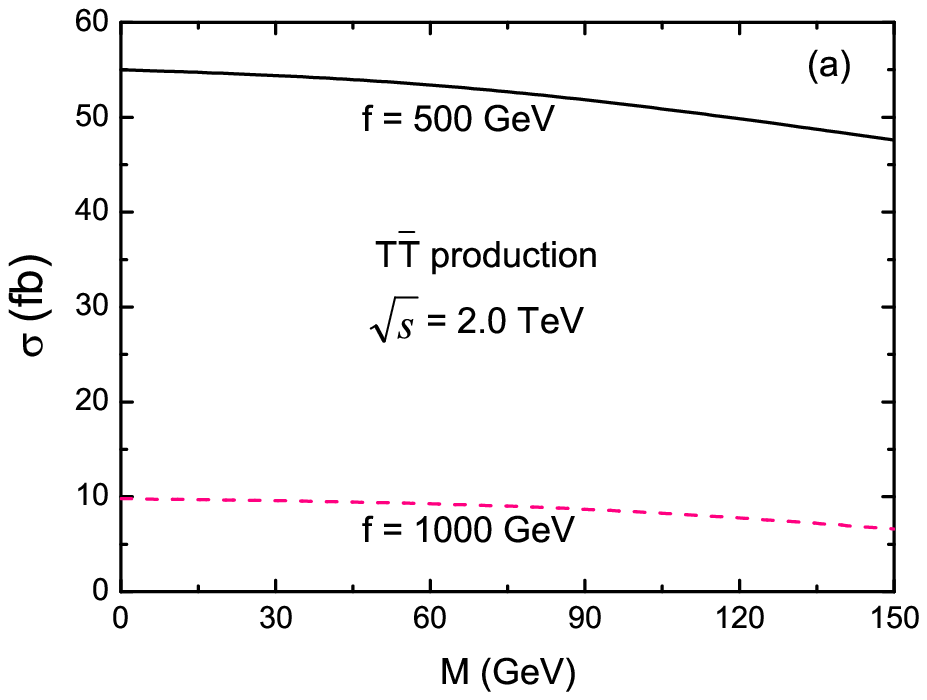}\epsfig{file=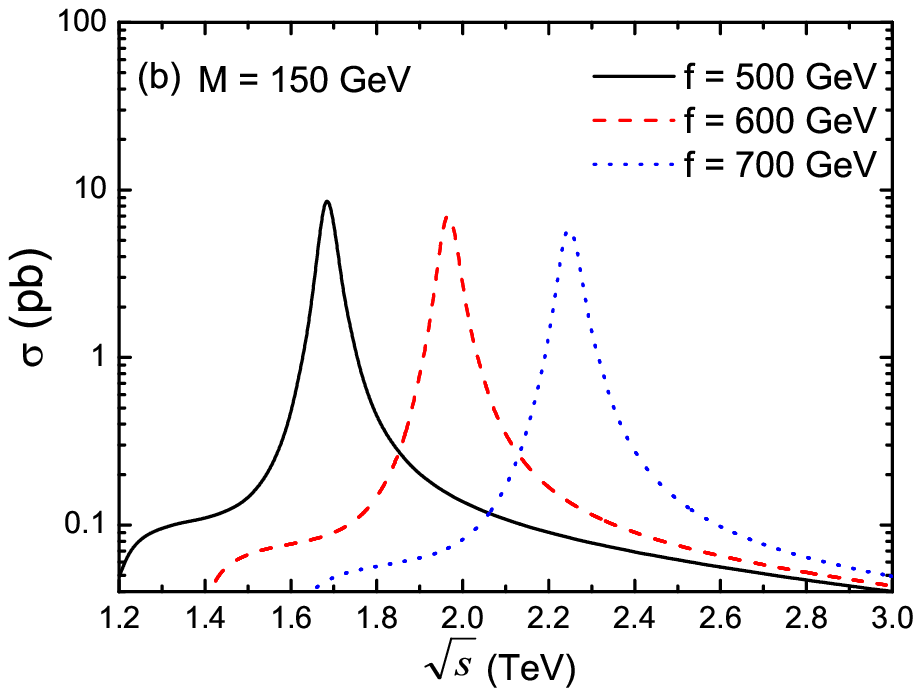}}
\end{center}
\caption{(a) The production CS $\sigma$ as a function
of the mixing parameter $M$ for $\sqrt{s}=2.0$ TeV and $f=500$GeV and 1000 GeV;
(b) The production CS $\sigma$ as a function
of center of mass energy $\sqrt{s}$ for $M=150$ GeV and three values of $f$ as indicated.}
\label{fig:fig7}
\end{figure}
%%=====================================

We next consider the pair production of the top partner $T$ at the CLIC. The production CS's
$\sigma$ are plotted as a function of the mixing parameter $M$ in Fig.~7a and as a function of $\sqrt{s}$
in Fig.~7b for various typical values of $f$.
From Fig.~7a, one can see that the cross section is insensitive to the parameter $M$.
For $f=500$ GeV, for example, the cross section $\sigma$ is changing from $55$ fb to $48$ fb when the parameter $M$
increases from 0 to 150 GeV. In the most of the parameter spaces, the production CS are at the
level of tens of fb for $\sqrt{s}=2.0$ TeV.
However, one can see from Fig.~7b that the resonance peak of the $\sigma$ can reach a few pb
when $M_{Z_{H}}\simeq \sqrt{s}$, provided that the LHC measures the masses
of the extra gauge bosons predicted by the LRTHM.
For $\sqrt{s}=3$ TeV this resonance scan can be extended to upper values of the scale $f$ around 1.1 TeV.

Considering the subsequent decay of the top partner $T$, the characteristic signal of $T\ov{T}$ events
might be:
\begin{itemize}
\item {\bf Case I}:
One lepton ($e$ or $\mu)$ + two jets +6b + missing $\eslash$, which arises from
the decay modes $\phi^{+}b$, $ht$, $Zt$, and $\phi^{0}t$ of the top partner $T$ with
the cascade decays $\phi^{+}\to t\bar{b}$, $t\to W^{+}b$, $h\to b\bar{b}$,
$Z\to b\bar{b}$ and $\phi^{0}\to b\bar{b}$,  and the subsequent decay of one $W$ bosons through leptonic decay channel
and others in their hadronic decays.

\item {\bf Case II}: Four jets +$2b$, which happens for a very small value of $M$
with $T\to \phi^{+}b$ and $\phi^{+}\to c\bar{s}$, eg., $M=0$.
\end{itemize}

%%=====================================
\begin{figure}[thb]
\begin{center}
\scalebox{0.75}{\epsfig{file=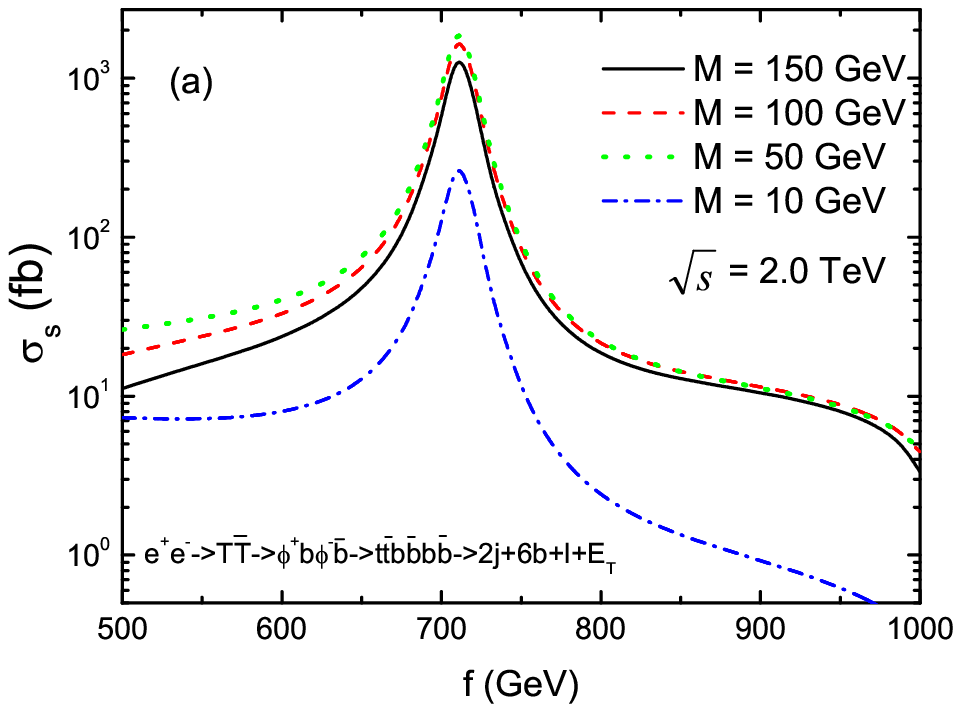}\epsfig{file=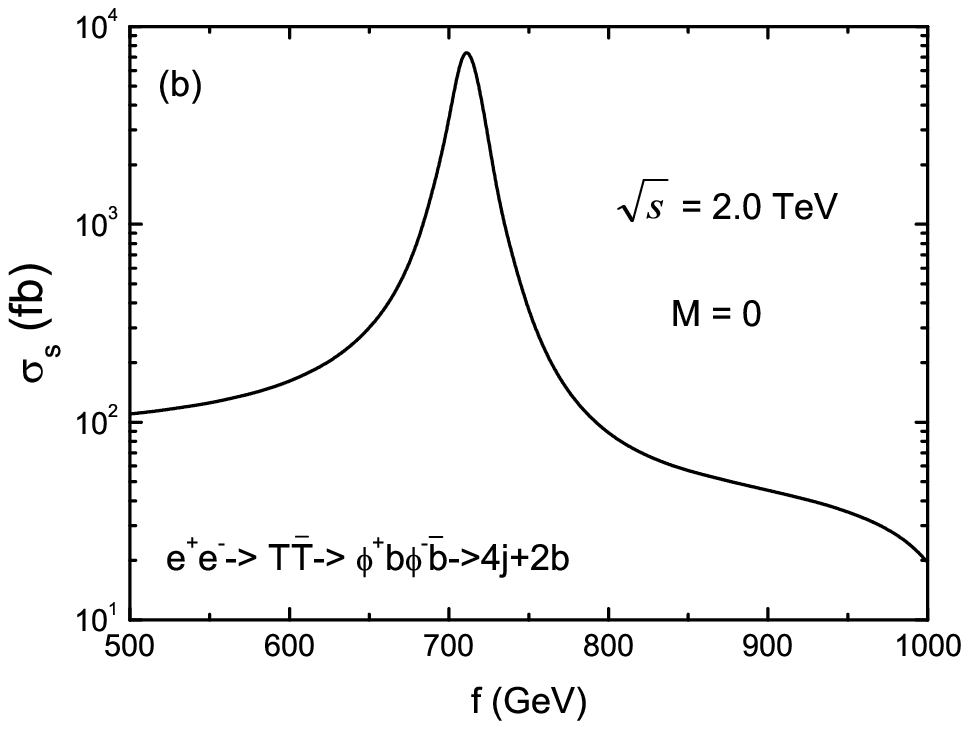}}
\end{center}
\caption{(a) The production rates of the $2j+6b+\ell+\eslash$ final state as a function
of  $f$ for $\sqrt{s}=2.0$ TeV and four typical values of $M$ as indicated;
(b) The production rates of the $4j+2b$ final state as a function of $f$ for $\sqrt{s}=2.0$ TeV and $M=0$.}
\end{figure}
%%=====================================

For $\sqrt{s}=2.0$ TeV, the total production rates of the signals for above two cases are shown in Fig. 8.
For Case I, the production rate of the signal can reach tens of fb except for the resonance effect, as shown in Fig.~8a.
While for case II, the production rate of the signal are higher about one order than that for Case I with the same value of parameter $f$, as shown in Fig.~8b.
For $f=600$ GeV and $M=0, 150$ GeV, the production rates for two cases are about 161 fb and 24 fb, respectively. If we assume that the future
CLIC experiment with $\sqrt{s}$=2.0 TeV has a yearly integrated
luminosity of 500fb$^{-1}$, then there will be about $10^{4}$ signal events generated per year.

For above two kinds of signals the possible backgrounds from the SM processes are listed in Table \ref{tab:table3}.
For Case I, one can see that the background are much smaller than the signal.
With the signal CS and the expected CLIC high luminosity, one can easily get large number of
events even if we lose some of events by imposing cuts to remove SM backgrounds.

\begin{table}[]
\begin{center}
\caption{ The SM background CS's (in fb) for $2j+6b+\ell+\eslash$ (Case I)
and $4j+2b$ (Case II) final states, estimated  with $\sqrt{s}=2.0$ TeV. } \label{tab:table3}
\vspace{0.1in}
\doublerulesep 0.8pt \tabcolsep 0.1in
\begin{tabular}{|l|l|}
\hline \multicolumn{2}{|c|}{Case I}\\
 \hline $\sigma(e^{+}e^{-}\to t\bar{t}Zh)=0.04$&$\sigma(e^{+}e^{-}\to t\bar{t}ZZ\to 2j+6b+l+\eslash)=9.8\times 10^{-4}$\\
\hline $\sigma(e^{+}e^{-}\to t\bar{t}hh)=0.011$&$\sigma(e^{+}e^{-}\to t\bar{t}hh\to 2j+6b+l+\eslash)=1.02\times 10^{-3}$\\
 \hline $\sigma(e^{+}e^{-}\to t\bar{t}ZZ)=0.056$&$\sigma(e^{+}e^{-}\to t\bar{t}ZZ\to 2j+6b+l+\eslash)=3.7\times 10^{-4}$\\
 \hline \multicolumn{2}{|c|}{Case II}\\
\hline $\sigma(e^{+}e^{-}\to t\bar{t})=43.8$&$\sigma(e^{+}e^{-}\to t\bar{t}\to W^{+}bW^{-}\bar{b}\to 4j+2b)=20.3$\\
\hline $\sigma(e^{+}e^{-}\to W^{+}W^{-}Z)=43.4$&$\sigma(e^{+}e^{-}\to W^{+}W^{-}Z\to 4j+2b)=3.01$\\
\hline $\sigma(e^{+}e^{-}\to W^{+}W^{-}h)=1.8$&$\sigma(e^{+}e^{-}\to W^{+}W^{-}h\to 4j+2b)=0.47$\\
\hline $\sigma(e^{+}e^{-}\to ZZb\bar{b})=0.19$&$\sigma(e^{+}e^{-}\to ZZb\bar{b}\to 4j+2b)=0.09$\\
\hline $\sigma(e^{+}e^{-}\to ZZh)=0.13$&$\sigma(e^{+}e^{-}\to ZZh\to 4j+2b)=0.04$\\
\hline $\sigma(e^{+}e^{-}\to ZZZ)=0.5$&$\sigma(e^{+}e^{-}\to ZZZ\to 4j+2b)=0.04$\\
\hline
 \end{tabular}\end{center}\end{table}

For Case II, the large background comes from the SM process $e^{+}e^{-}\to t\bar{t}\to 2W+2b\to 4j+2b$ with the
cross section about 20 fb.
Since the cross sections of the SM processes are not too large compared to the signal process, the
reconstruction of top partner $T$ and the charged Higgs bosons $\phi^+$ is necessary to distinguish the
signal from the background. For example, one must first search for the hadronic decay
of a charged Higgs boson by choosing the combination which minimizes
$|m_{jj}-m_{\phi}|$. An apparent feature of difference between signal and the
background is that the di-jet invariant mass for the background
events primarily  reconstructs to $m_{W}$ but the di-jet invariant mass for the signals
coming from the charged Higgs  approaches $m_{\phi}$. Such difference can
help us to distinguish the signals from the background.
Secondly, each top partner $T$ is reconstructed from one charged Higgs candidate paired with one of the two $b$
jets, such that the invariant masses of the $\phi^{+} b$ systems are as close as possible to top partner mass.

%%============================================
\begin{figure}[thb]
\begin{center}
\scalebox{0.75}{\epsfig{file=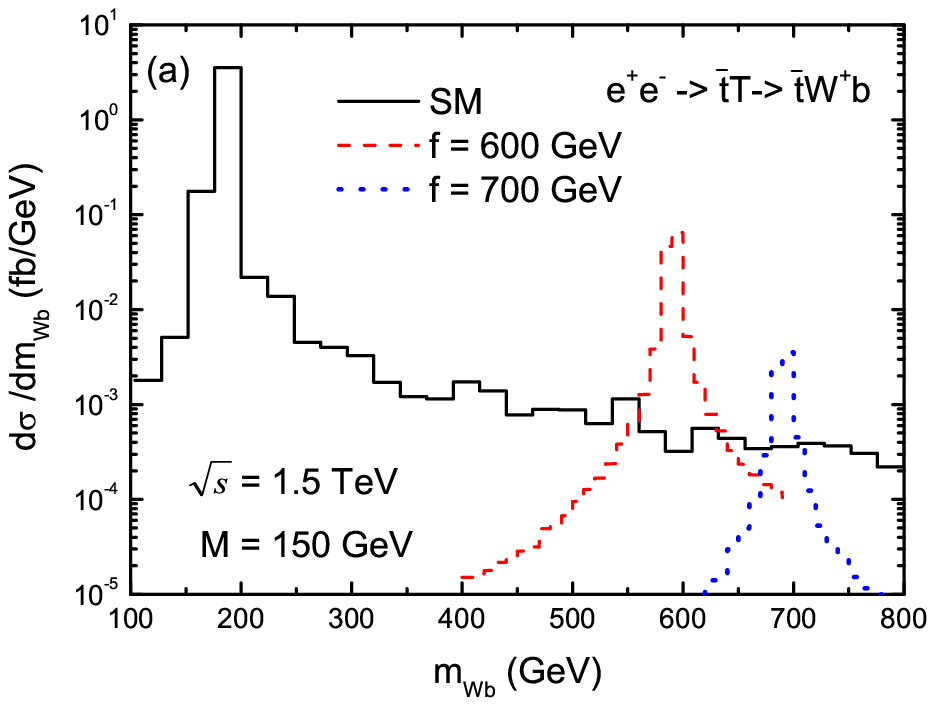}\epsfig{file=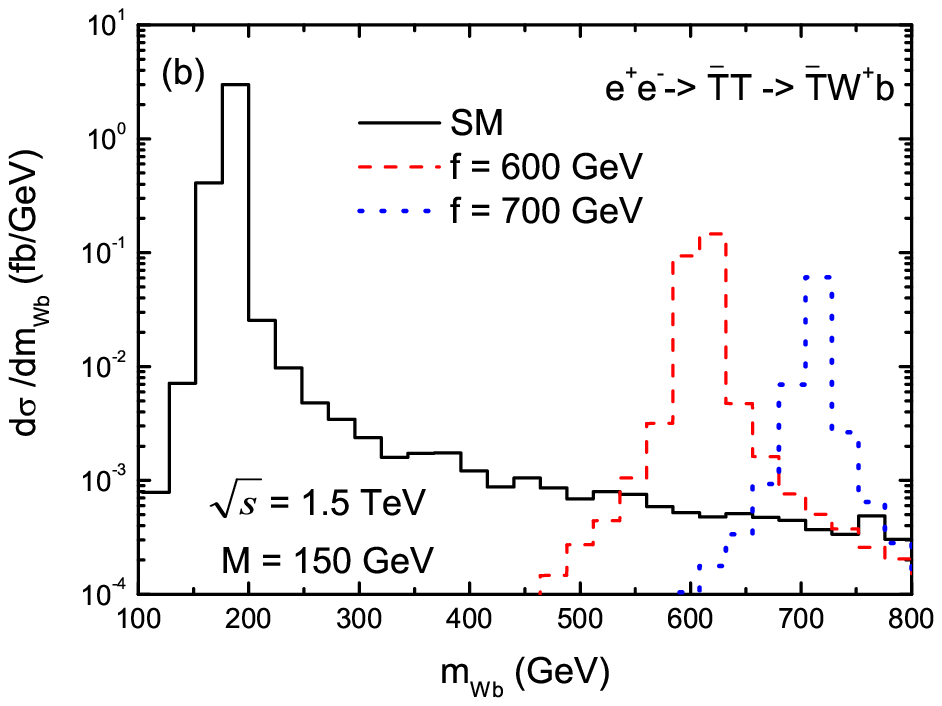}}
\end{center}
\caption{ The invariant mass distributions for the SM
background and the $Wb$ signal from $T$ decay for
(a) $e^{+}e^{-}\to \bar{t}T\to \bar{t}W^{+}b$ and (b) $e^{+}e^{-}\to \ov{T}T\to \ov{T}W^{+}b$. }
\label{fig:fig9}
\end{figure}
%%============================================

For the decay channel $T\to W^{+}b$, the top partner production can give rise to the same final state as the SM top quark.
The leptonic $W$ decay yields a nice signal of one $b$ jet plus one electron or muon with missing energy.
For $M=150$ GeV and $f=600, 700$ GeV, the branching ratios of $T\to W^{+}b$ are about $7.8\%$ and $5.5\%$, respectively.
The invariant mass distributions for the SM
background and the $Wb$ signal from $T$ decay are  shown in Fig.~9 for two processes with
$\sqrt{s}=1.5$ TeV and $M=150$ GeV. It is clear that the $T$-quark signal can be observed as a resonance in
the $W^{+}b$ invariant mass distribution at the CLIC.

\subsection{Associate productions of $T$  with SM-like Higgs boson $h$}

Like $ht\bar{t}$ production, the productions of $ht\ov{T}$ can also be realized at the linear $e^{+}e^{-}$ collider,
as shown in Fig.~10. Thus, it is possible to measure the Yukawa
coupling between top partner and other particles simply by measuring the production CS's of
the relevant processes with high center of mass energy.
There are two SM-like Higgs boson associated production processes. One is the Higgs production associating with a
top quark and a top partner production $e^{+}e^{-}\to t\ov{T}h (T\bar{t}h)$, and another is the process associating
with top partner pairs $e^{+}e^{-}\to T\ov{T}h$. Here we fixed the SM-like Higgs boson mass as $m_{h}=125.5$ GeV.
Considering the dominant decay mode $T\to \phi^{+}b\to t\bar{b}b$, the $t\ov{T}h$ and
 $T\ov{T}h$ production processes have less background than $ht\bar{t}$ production and these new production channels
 at the LHC have been studied in \cite{sjf}.

%%=================================================================
\begin{figure}[thb]
\begin{center}
\epsfig{file=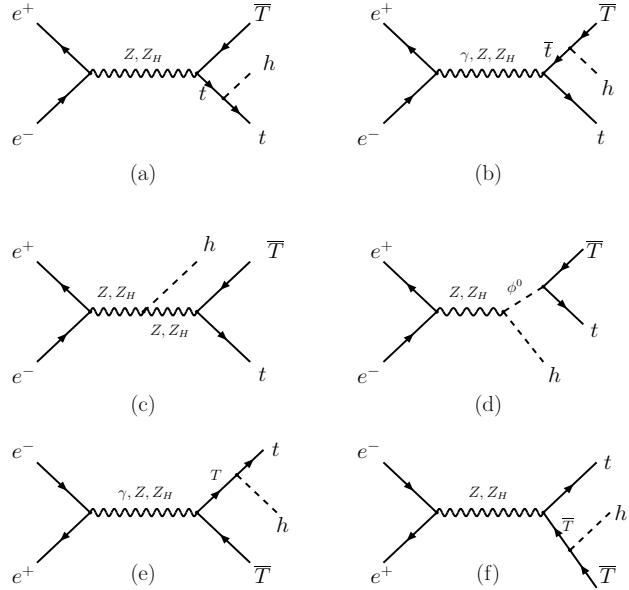,width=300pt,height=400pt}
\vspace{-5cm}
\caption{ Typical Feynman diagrams of the process $e^{+}e^{-}\to t\ov{T}h$ in the LRTHM.}
\label{fig:fig10}
\end{center}
\end{figure}
%%=================================================================

\subsubsection{The $e^{+}e^{-}\to t\ov{T}h+T\bar{t}h$ process}

We first consider the Higgs production process associated with
a top quark and a top partner. The sum of the CS, $\sigma(e^{+}e^{-}\to t\ov{T}h)+\sigma(e^{+}e^{-}\to T\bar{t}h)$,
are shown in Fig.~11.
One can see that in the major region of the parameter space, the CS are at the level of several fb for $M=150$ GeV.
For example, the CS is about 3.6 fb for $\sqrt{s}=1.5$ TeV and $f=700$ GeV.
On the other hand, the resonance peak values of the
$\sigma$ can reach the order of $10^{2}$ fb. The production CS is, furthermore,
very sensitive to the parameter $M$: large values of $M$ can enhance the CS significantly.
This is due to the couplings of $t\ov{T}h$, $Zt\ov{T}$ and $Z_{H}t\ov{T}$ are all proportional to the
factor $(M/f)$. In the limit of $M=0$, all theses couplings are vanishing.
For $\sqrt{s}=3$ TeV, $f$=1200 GeV, the value of $\sigma$ is changing from $0.02$ fb to $0.42$ fb
when the parameter $M$ increases from 30 GeV to 150 GeV.

%%==========================================================
\begin{figure}[thb]
\begin{center}
\vspace{-0.5cm}
\centerline{\epsfxsize=8cm\epsffile{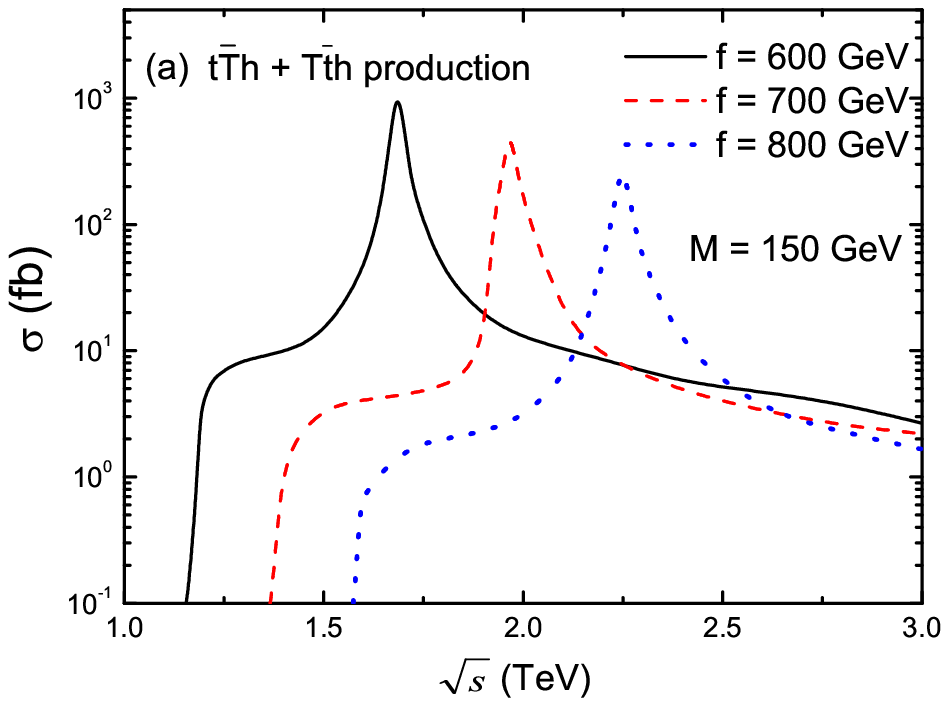}
\hspace{-0.5cm}\epsfxsize=8cm\epsffile{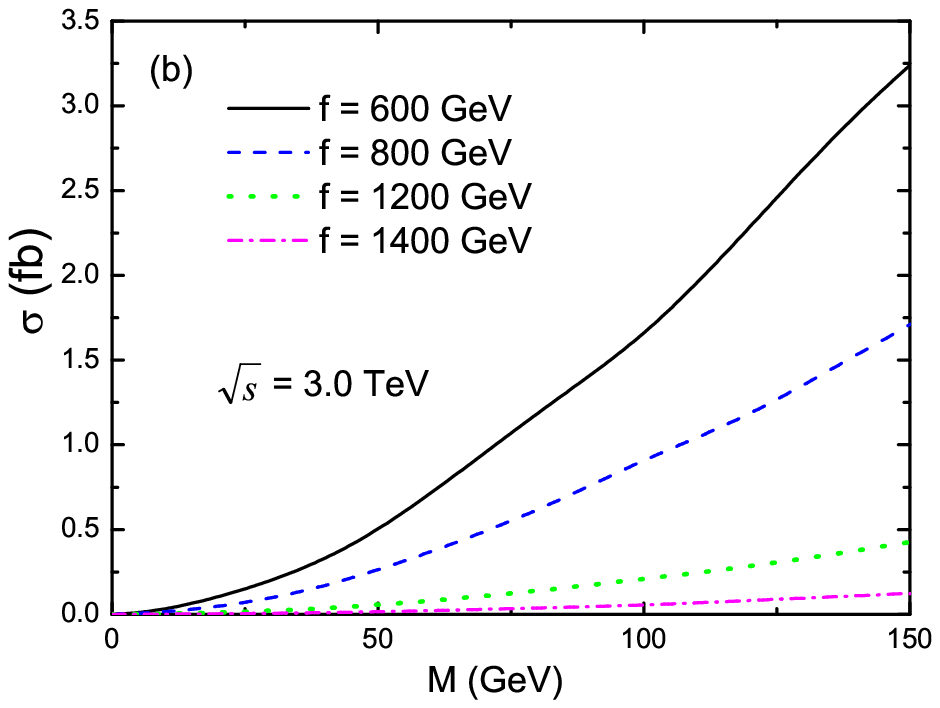}}
\caption{(a) The production CS $\sigma$ as a function
of $\sqrt{s}$ for $M=150$ GeV and three values of $f$ as indicated;
(b) The production CS $\sigma$ as a function
of the mixing parameter $M$ for $\sqrt{s}=3.0$ TeV and $f=600, 800, 1200, 1400$ GeV respectively.}
\label{fig:fig11}
\end{center}
\end{figure}
%%==========================================================

For a large value of $M$, the dominate decay mode $h\to b\bar{b}$ will lead to the cascade decay chain
$e^{+}e^{-}\to t\ov{T}h+T\bar{t}h \to t\bar{t}b\bar{b}b\bar{b}$.
The production rates for the final state $t\bar{t}b\bar{b}b\bar{b}$
can be easily estimated:
\beq
\sigma_{s}&\simeq & \sigma\times \Bigl [ Br(T\to \phi^{+}b)\cdot Br(\phi^{+}\to t\bar{b})
+Br(T\to th)\cdot Br(h\to b\bar{b}) \non
&& +Br(T\to tZ)\cdot BR(Z\to b\bar{b})+ Br(T\to t\phi^{0})\cdot Br(\phi^{0}\to b\bar{b})\Bigr ].
\eeq
In Table \ref{tab:table4} we present the total CS for the final states $t\bar{t}b\bar{b}b\bar{b}$
via the process $e^{+}e^{-}\to t\ov{T}h+T\bar{t}h$ with $\sqrt{s}=3.0$ TeV and various parameter values.
The main backgrounds for the
$t\bar{t}b\bar{b}b\bar{b}$ final state come from the SM processes
$e^{+}e^{-}\to t\bar{t}ZZ$, $e^{+}e^{-}\to
t\bar{t}Zh$ and $e^{+}e^{-}\to t\bar{t}hh$ with
$Z\to b\bar{b}$ and $h\to b\bar{b}$, continuum
$t\bar{t}b\bar{b}b\bar{b}$ production.
The total CS of the SM backgrounds is estimated about 0.01 fb, which is smaller than that in the signal.
Thus, it may be possible to extract the signals from the backgrounds in the reasonable parameter
spaces in the LRTHM (eg., for large $M$ and small $f$).

\begin{table}[]
\begin{center}
\caption{ The total CS's (in fb) of signal
for the final states $t\bar{t}b\bar{b}b\bar{b}$ with $\sqrt{s}=3.0$ TeV.}
\label{tab:table4}
\vspace{0.1in}
\doublerulesep 0.8pt \tabcolsep 0.1in
\begin{tabular}{|l|l|l|l|l|} \hline $f$ (GeV)&600&800&1200&1400  \\ \hline
$M=50$ GeV&0.42 &0.23&0.047 &0.013
\\ \hline
$M=100$ GeV&1.43 &0.87 &0.21 &0.053
\\ \hline
$M=150$ GeV&2.56 &1.53 &0.4 & 0.12 \\
\hline
\end{tabular} \end{center}\end{table}
\vspace{0.2cm}

\subsubsection{The $e^{+}e^{-}\to T\ov{T}h$ process}

%%===========================================================
\begin{figure}[thb]
\begin{center}
\vspace{-0.5cm}
\centerline{\epsfxsize=8cm\epsffile{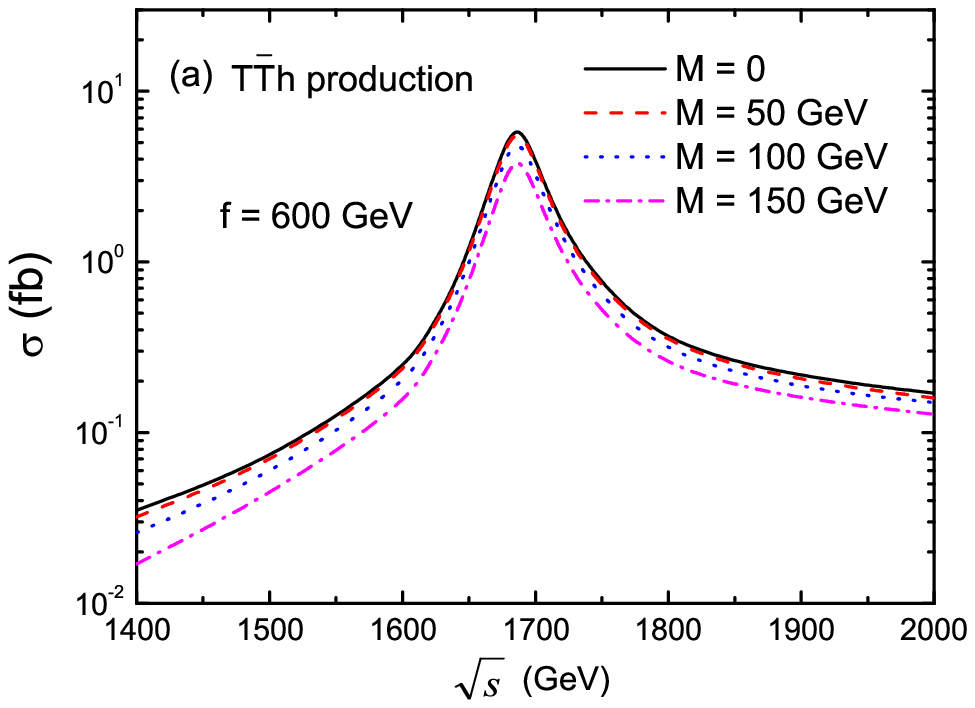}
\hspace{-0.5cm}\epsfxsize=8cm\epsffile{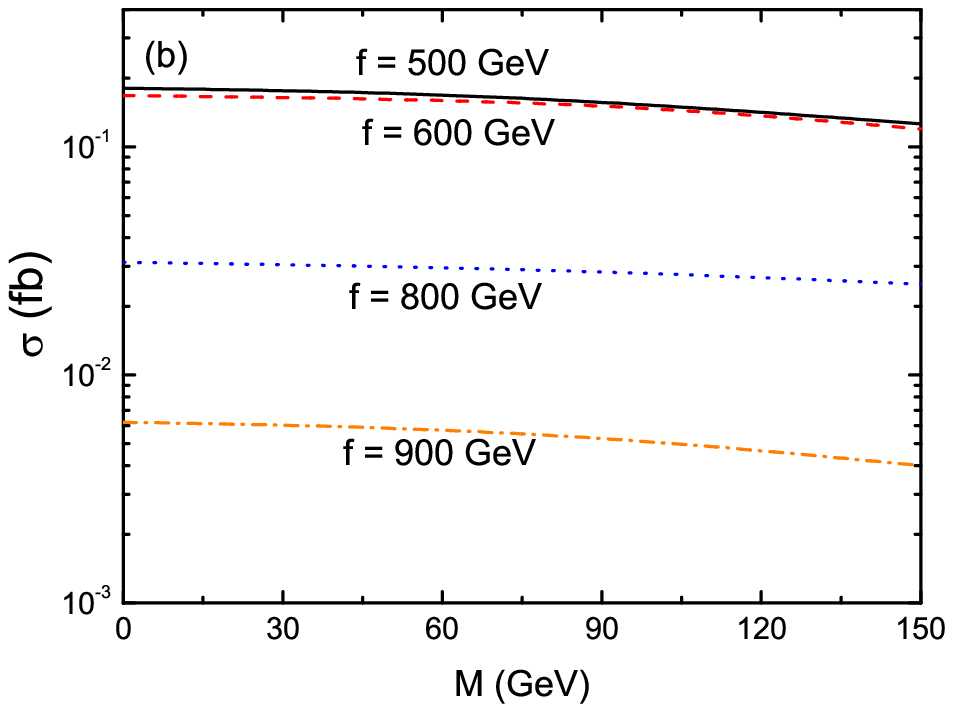}}
\caption{(a) The production CS $\sigma$ as a function
of $\sqrt{s}$ for four values of $M$ as indicated and $f=700$ GeV;
(b)The production CS $\sigma$ as a function
of the mixing parameter $M$ for $\sqrt{s}=2.0$ TeV and four typical values of $f$ as indicated.}
\label{fig:fig12}
\end{center}
\end{figure}
%%============================================================

Next, we consider the pair production of the top partner $T$ associated with the Higgs boson:
$e^{+}e^{-}\to T\ov{T}h$.
In Fig.~12a, we show its production CS versus $\sqrt{s}$ with various $M$ for
$f=700$ GeV. One can see that the resonance CS's are at the level of several fb.
In the most parameter space, the CS's are smaller than 0.1 fb.
From Fig.~12b one can see that the CS decrease along with the increase of $f$,
and is also insensitive to the variation of $M$.
For $f=800$ GeV, the CS $\sigma$ is changing from $0.03$ fb to $0.025$ fb when the parameter $M$
increases from 0 to 150 GeV.

Similar to the character of $e^{+}e^{-}\to T\ov{T}$ process, the characteristic signal of $T\ov{T}h$
with $h\to b\bar{b}$ might be
\begin{itemize}
\item {Case A}:
$2j+8b+\ell+\eslash$ for $M=150$ GeV, which arises from the semi-leptonic decays of the $t\bar{t}$ system.

\item {Case B}:
$4j+4b$ in the limit of $M=0$, which arises from $T\to \phi^{+}b$ and $\phi^{+}\to c\bar{s}$
with the branching ratios of $100\%$.
\end{itemize}

%%------------------------------------------------------------------------------------------------
\begin{table}[]
\begin{center}
\caption{ The possible signal cross sections (in fb) for above two cases are estimated with
$\sqrt{s}=2.0$ TeV.}\label{tab:table5}
\vspace{0.1in} \doublerulesep 0.8pt \tabcolsep 0.1in
\begin{tabular}{|l|l|l|l|} \hline
Signals&$f=600$ GeV&$f=800$ GeV&$f=900$ GeV\\ \hline
Case A&$1.4\times 10^{-2}$&$3.3\times 10^{-3}$&$5.6\times 10^{-4}$\\ \hline
Case B&$9.6\times 10^{-2}$&$1.8\times 10^{-2}$&$3.5\times 10^{-3}$\\ \hline
\end{tabular}\end{center}\end{table}
%%------------------------------------------------------------------------------------------------

The CS's of possible signals are listed in Table \ref{tab:table5} with $\sqrt{s}=2.0$ TeV.
The reducible SM backgrounds for Case A are almost negligible. Given a sufficient integrated
luminosity, it may be possible to detect these signals in the reasonable parameter space of
the LRTH model, especially for small value of $f$. The main background processes for the Case B
have been extensively studied in \cite{tth1,tth2} by applying the suitable cuts.
According their conclusions, we have to say that it is very difficult to discriminate the
$4j+4b$ signal due to the low production rates, low selection efficiencies and large SM background.

\subsection{Associate production with $\phi^{0}$}

%%------------------------------------------------------------------
\begin{figure}[thb]
\begin{center}
\epsfig{file=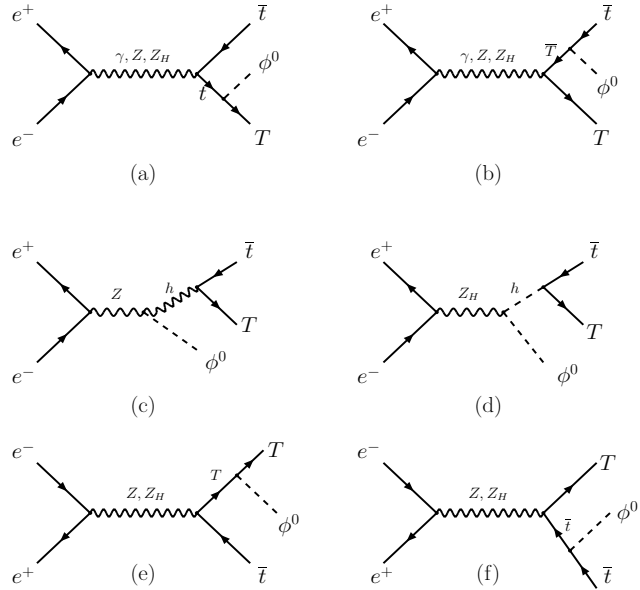,width=300pt,height=400pt}
\vspace{-5cm}
\caption{Feynman diagrams of the process $e^{+}e^{-}\to T\bar{t}\phi^{0}$ in the LRTHM.} \label{fig:fig13}
\end{center}
\end{figure}
%%------------------------------------------------------------------

Besides the SM-like Higgs boson $h$, the LRTHM also predicts the existence of the neutral pseudoscalar boson
$\phi^0$. In Ref.~\cite{liu-jhep}, we studied the production and decays of a light $\phi^0$.
The relevant couplings can be written as \cite{Hock}:
\begin{eqnarray}
\phi^{0}\ov{T}t&:&-iy(S_{L}C_{R}P_{L}-C_{L}S_{R}P_{R})/\sqrt{2},\non
\phi^{0}T\ov{T}&:&-iyC_{L}C_{R}\gamma_{5}/\sqrt{2},\non
h\phi^{0}Z_{\mu}&:&iexp3_{\mu}/(6s_{W}c_{W}),\non
h\phi^{0}Z_{H\mu}&:&iex[(14-17s_{W}^{2})p2_{\mu}-(4-s_{W}^{2})p1_{\mu}]/(18s_{W}c_{W}\sqrt{1-2s_{W}^{2}}),
\end{eqnarray}
where $P_{L,R}=(1\mp\gamma_{5})/2$, $p1$, $p2$ and $p3$ refer to the incoming momentum
of the first, second and third particle, respectively.
It is easy to see that the top partner $T$ can be produced via the process $e^{+}e^{-}\to  \phi^{0}T\bar{t}$,
as shown in Fig.~13. Similarly, the associated production of $\phi^{0}t\ov{T}$ and $\phi^{0}T\ov{T}$ can also
happen although we do not show them explicitly in Fig.~13.

\subsubsection{The $e^{+}e^{-}\to t\ov{T}\phi^{0}+T\bar{t}\phi^{0}$ process}

In Fig.~14, we plot the parameter dependence of the summation of the production CS,
$\sigma(e^{+}e^{-}\to t\ov{T}\phi^{0})+\sigma(e^{+}e^{-}\to T\bar{t}\phi^{0})$.
This case is similar to those in the SM-like Higgs boson associate production processes.
One can see that in the considered parameter space, the production CS are at the level of
several fb for $M=150$ GeV.
The resonance peak values of the $\sigma$ can reach the order of $10^{2}$ fb.
On the other hand, the production CS is very sensitive to the
parameter $M$ and decreases along with the increase of $m_{\phi^0}$.
For $\sqrt{s}=2.0$ TeV, $f$=600 GeV and $m_{\phi^{0}}=120$ GeV, the value of
$\sigma$ is changing from $0.26$ fb to $4.2$ fb when the parameter $M$
increases from 30 GeV to 150 GeV.  Thus, a large value of $M$ can enhance the production rates for this process.

%%==========================================================
\begin{figure}[thb]
\begin{center}
\vspace{-0.5cm}
\centerline{\epsfxsize=8.5cm\epsffile{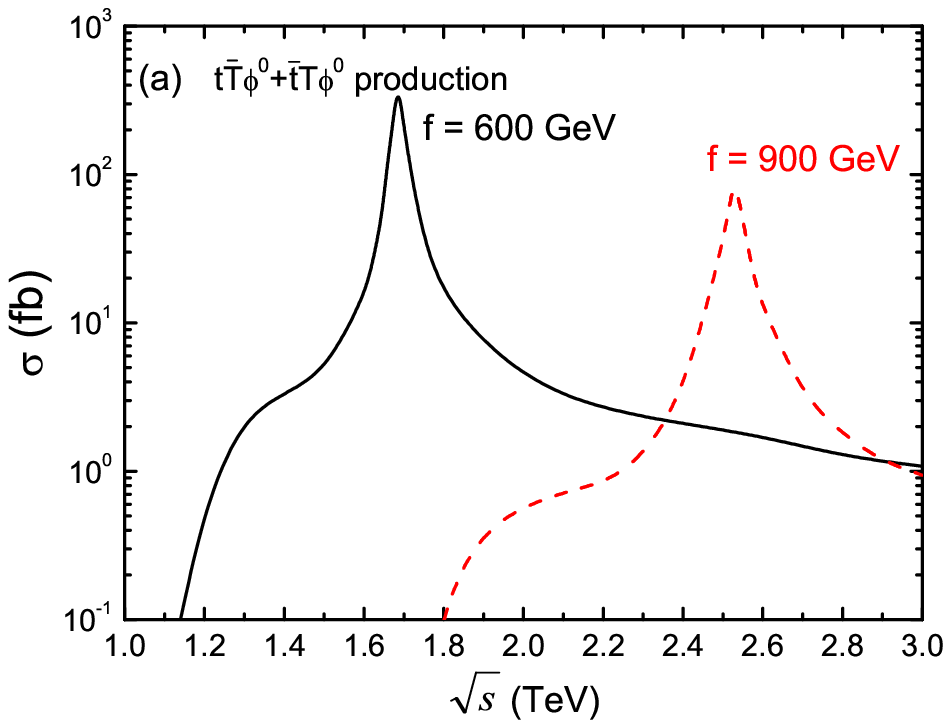}
\hspace{-0.5cm}\epsfxsize=8.5cm\epsffile{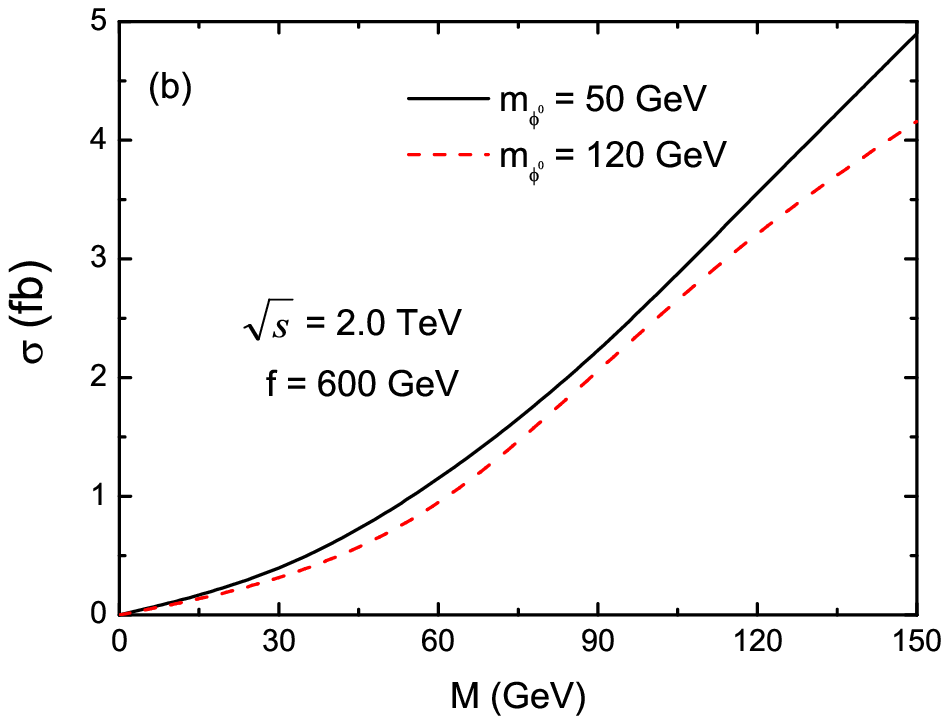}}
\caption{(a) The production CS as a function
of $\sqrt{s}$ for $M=150$ GeV, $m_{\phi^{0}}=120$ GeV and $f=600, 900$ GeV;
(b) The production CS $\sigma$ as a function
of the mixing parameter $M$ for $\sqrt{s}=2.0$ TeV, $f=600$ GeV, and $m_{\phi^{0}}=50, 120$ GeV.}
\label{fig:fig14}
\end{center}
\end{figure}
%%==========================================================

The dominant decay mode of $\phi^0$ is $\phi^0\to b\bar{b}$, with a branching ratio $Br(\phi^0\to b\bar{b})\simeq
0.8$ fb for $m_{\phi^{0}}=120$ GeV \cite{liu-jhep}.
For a large value of $M$, the dominate decay mode $T\to \phi^+b\to t\bar{b}b$ can also make the process
$e^{+}e^{-}\to t\ov{T}\phi^0+T\bar{t}\phi^0$ also give rise to the
$t\bar{t}b\bar{b}b\bar{b}$ final state, which is similar to the case of $(t\ov{T}h+T\bar{t}h)$ productions.
The branching ratio for the $t\to bW^{+}$ is essentially one which induced to the final state of
$6b+2W$. Now we consider one $W$ boson decay hadronically and the other decay leptonically.
Thus the resulting final state signal is $2j+6b+\ell+\eslash$.
The production rates of such final state are shown in Table \ref{tab:table6}
with $m_{\phi^{0}}=120$ GeV, $\sqrt{s}=2.0$ TeV and various parameter values.
For $f=600$ GeV and $M=150$ GeV, there will be about 230 signal events with a yearly integrated
luminosity of 500fb$^{-1}$.
The relevant SM backgrounds for this final state is negligible.
Note that what we have presented here as an estimate of the signal events is just a rude estimate.
If we take the $b$-tagging efficiency of each of the six $b$ quarks which is about $70\%$, the estimated
event rates are suppressed
about $(0.7)^{6}\simeq 0.12$ and this still gives us tens of observable events for the signal with high luminosity.
Thus, it may be possible to extract the signals from the backgrounds due to the large production rates
in the reasonable parameter spaces of the LRTHM.

%%------------------------------------------------------------------------------
\begin{table}[]
\begin{center}
\caption{ The total cross sections (in fb) of signal for the final states $2j+6b+\ell+\eslash$ in the
LRTHM for $m_{\phi^{0}}=120$ GeV and $\sqrt{s}=2.0$ TeV.}\label{tab:table6}
\vspace{0.1in} \doublerulesep 0.8pt \tabcolsep 0.1in
\begin{tabular}{|c|c|c|c|c|c|} \hline $M$ (GeV)&30&60&90&120&150  \\ \hline
f=600 GeV&0.03 &0.11&0.26 &0.39&0.46 \\ \hline
f=900 GeV&0.003 &0.012&0.028 &0.045&0.074 \\ \hline
\end{tabular} \end{center}\end{table}
%%-------------------------------------------------------------------------------

\subsubsection{The $e^{+}e^{-}\to T\ov{T}\phi^{0}$ process}

%%===================================================================
\begin{figure}[thb]
\begin{center}
\vspace{-0.5cm}
\centerline{\epsfxsize=8.5cm\epsffile{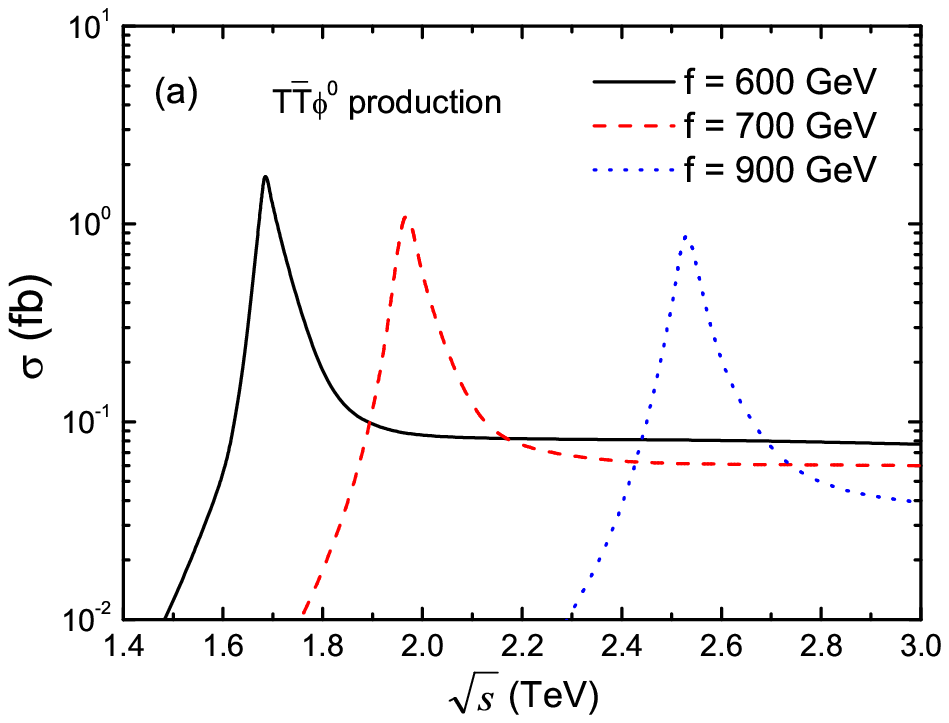}
\hspace{-0.5cm}\epsfxsize=8.5cm\epsffile{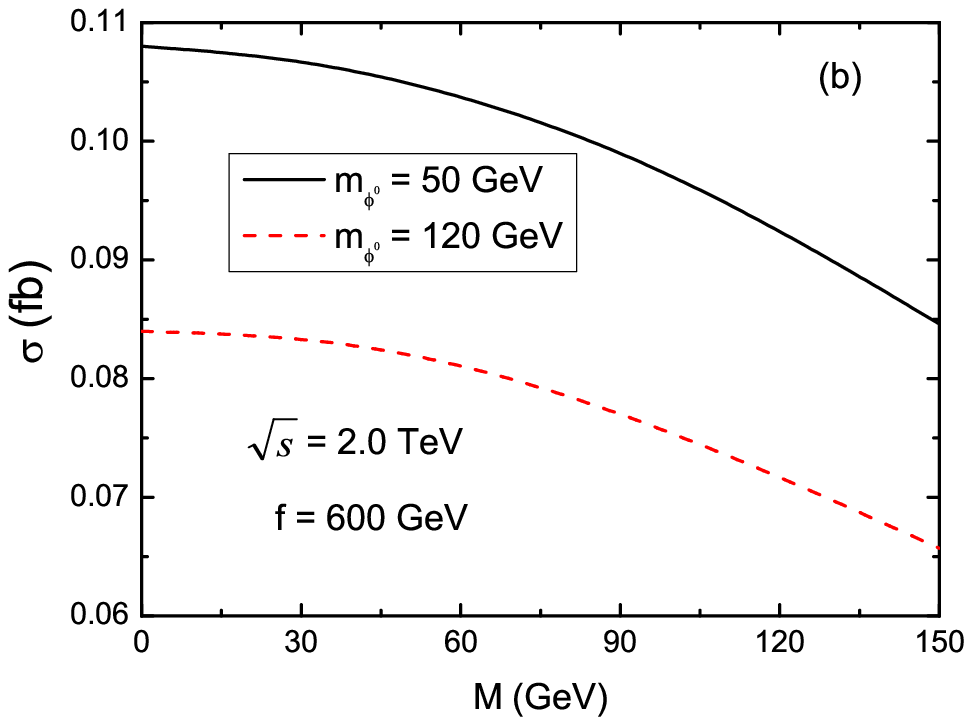}}
\caption{(a) The production CS $\sigma$ as a function
of $\sqrt{s}$ for $M=150$ GeV, $m_{\phi^{0}}=120$ GeV and three values of $f$ as indicated; (b)The production CS $\sigma$ as a function
of the mixing parameter $M$ for $\sqrt{s}=2.0$ TeV, $f=600$ GeV, and $m_{\phi^{0}}=50, 120$ GeV.}
\end{center} \end{figure}
%%====================================================================

The production CS of the process $e^{+}e^{-}\to T\ov{T}\phi^{0}$ are shown in Fig.~15. One can see that the resonance CS's
can reach the level of 1 fb. Apart from the resonance peak, the cross sections are smaller than 0.1 fb in the most
parameter space.
For $f=600$ GeV, $m_{\phi^{0}}=120$ GeV and $\sqrt{s}=2.0$ TeV, the cross section $\sqrt{s}$ is
changing from $0.084$ fb to $0.066$ fb when the parameter $M$
increases from 0 to 150 GeV. Thus, it is challenging to detect the signals of top partner via
this production process due to the small production rates, except for the resonant region.

\section{Conclusions}

The LRTHM predicts the existence of the top partner $T$ which may be observable at the high energy linear
$e^{+}e^{-} $colliders.
In this paper, we study the single and pair production of the top partner at the
ILC and CLIC via the processes: $e^{+}e^{-}\to (t\ov{T},T\bar{t},T\ov{T})$, the Higgs boson $h$ associate
productions $e^{+}e^{-}\to (t\ov{T}h, T\bar{t}h, T\ov{T}h)$, and the neutral pseudoscalar boson associate
productions $e^{+}e^{-}\to (t\ov{T}\phi^{0}, T\bar{t}\phi^{0})$ and
$e^{+}e^{-}\to T\ov{T}\phi^{0}$.
From the numerical calculations and the phenomenological analysis for all considered
production and decay modes, we find the following observations:
\begin{enumerate}
\item
The top partner $T$ mainly decay into $\phi^{+}b$ with the branching ratio larger than $60\%$ for
500 GeV $\leq f\leq$1000 GeV, while the branching ratio of $T\to Wb$ mode is about
$11\%$ for $M=150$ GeV and $f$=500 GeV. The current bound on the top partner mass $m_{T}$ could  be
relaxed.

\item
For the single top partner production processes: $T\bar{t}$, $T\bar{t}h$, and $T\bar{t}\phi^{0}$, the production CS's
are sensitive to the mixing parameter $M$, and will increase when the mixing parameter $M$ becomes larger.
Except for the resonance regions, the production CS's can reach the level of several fb for $M=150$ GeV.

\item
For the pair production process $e^{+}e^{-}\to T\ov{T}$, the production CS's are insensitive to the
mixing parameter $M$, and the production CS's can reach the level of tens of fb.
However, the production CS's of the processes $e^{+}e^{-}\to T\ov{T}h$ and $e^{+}e^{-}\to T\ov{T}\phi^{0}$
are smaller than 0.1 fb in the major part of the parameter space in the LRTHM.

\item
For the cases of the resonant production, the position and the shape of the peak of the production CS
have strong dependence of the value of the parameter $f$.
The subsequent decay of $T\to\phi^{+}b$, $\phi^{+}\to t\bar{b}$, $t\to W^{+}b$ and $W\to \ell \nu$
can give rise to the signal of the top partner $T$ with the $3b+\ell+ \eslash$, which can generate
typical phenomenological features for the top partners in the LRTHM.

\item
According to our SM background analysis, we get to know that the signal of the top partner $T$
predicted by the LRTHM, in the reasonable parameter space ( say small $f$ and large $M$),
may be detectable in the future ILC and CLIC experiments.
\end{enumerate}

\begin{acknowledgments}

We thank Shufang Su for providing the CalcHep Model Code. This work is supported by the National Natural Science Foundation of China under the Grant No. 11235005,  the Joint Funds
of the National Natural Science Foundation of China (U1304112) and by the Project on Graduate Students Education and Innovation of Jiangsu Province under Grant No. KYZZ-0210.

\end{acknowledgments}

%%%%%%%%%%%%%%%%%%%%%%%%%%%%%%%%%%%%%%%%%%%%%%%%%%%%%%%%%%%%%%%%%%%%%%%%%%%%%%%%%%%%%%%%%%%%%%5
%                                 reference
%%%%%%%%%%%%%%%%%%%%%%%%%%%%%%%%%%%%%%%%%%%%%%%%%%%%%%%%%%%%%%%%%%%%%%%%%%%%%%%%%%%%%%%%%%%%%%%%%

\end{document}